\documentstyle[12pt,aaspp4]{article}

\begin{document}

\title{CCD PHOTOMETRY OF GALACTIC GLOBULAR CLUSTERS. IV.  THE NGC 1851 RR LYRAES}

\author{Alistair R. Walker}
\affil{Cerro Tololo Inter-American Observatory, National Optical Astronomy
Observatories\altaffilmark{1}}

\authoraddr{Casilla 603, La Serena, Chile    \\}

\altaffiltext{1}{
Operated by the Association of Universities for Research in Astronomy, Inc.,
under cooperative agreement with the National Science Foundation.}

\begin{abstract}

The variable star population of the galactic globular cluster NGC 1851(C0512-400) 
has been studied by CCD photometry, from observations made in the B, V and I bands 
during 1993-1994.  Light curves are presented for 29 variables, seven of which are 
new discoveries.  The behavior of the RR Lyraes in the period-temperature diagram 
appears normal when compared to clusters which bracket the NGC
1851 metallicity.  Reddening and metallicity are re-evaluated, with no compelling 
evidence being found to change from the values of $E(B-V) = 0.02$ and $[Fe/H] = -1.29$
(Zinn scale) adopted in recent studies of the cluster.  Photometry is provided for stars 
in an annulus with radii 80 and 260 arcsec centered on NGC 1851.  To at least $V=18.5$
there is excellent agreement with the extensive earlier photometry for the brighter
NGC 1851 stars, with systematics less than 0.02 mag in all colors.  Instability strip
boundary positions for several clusters shows a trend for the red boundary to move to 
redder colors as the metallicity increases.   

keywords: globular clusters: individual (NGC1851) - RR Lyrae variable

\end{abstract}

\section{Introduction}

The galactic globular cluster (GC) NGC 1851 (C0512-400) is rich, centrally-condensed 
and belongs to the small group of clusters which display bimodal horizontal-branch (HB)
morphology, defined (Catelan et al 1998) as having fewer RR Lyrae stars than either blue 
or red HB stars.   Canonical theory, that is considering a GC as a population characterized 
by a single age, constant abundance and a red-giant branch (RGB) mass loss parameter that 
has a narrow Gaussian distribution (typically $\sigma_M \sim 0.02 M_{\sun}$), cannot explain 
such unusual HB morphology.

In order to account for the bimodality, attention has focussed recently on scenarios which 
can alter the mass loss parameter, such as tidal stripping of red-giant envelopes in dense 
environments, rapid rotation, stellar encounters, and binary interactions.  Sosin et al 
(1997) discuss these various options in the context of the most extreme example known
of a GC with bimodal HB, NGC 2808, which displays a blue HB with multiple gaps that 
extends to below the main sequence turnoff in the $V, B-V$ color-magnitude diagram 
(CMD),  $M_V \sim 5$.  They conclude that none of the present explanations are a satisfactory 
match to the observations.   However, Sweigart \& Catelan (1997) have modeled the unusual 
HB morphology of the metal rich GC's 
NGC 6338 ($[Fe/H] = -0.60$) and NGC 6441 ($[Fe/H = -0.53$) for which Rich et al (1997) 
have obtained CMD's using the Hubble Space Telescope.  Both clusters have HB's which slope 
upwards (brighter) with decreasing $B-V$, and have extended blue tails.  Models with high helium
abundance, rapid rotation, and helium mixing into the envelope are all able to produce a 
sloping HB morphology, and sometimes a bimodal distribution.  The helium mixing alternative 
is particularly interesting given the observed heavy-element abundance variations in globular 
cluster red-giant stars (Kraft 1994).  Mixing deep enough to produce enhanced aluminium, as 
observed in some stars, will also dredge up helium.  Extensive deep mixing might be expected
to destroy the sharp boundary corresponding to the deepest penetration of the convective zone, 
and thus prevent the observational pile-up of stars on the RGB near the level of the HB.
NGC 1851 in fact appears to have quite a prominent such clump, thus suggesting that deep
mixing has not taken place on the RGB at a luminosity less than that of the clump.
Notwithstanding, the number of possible options still available to explain the peculiar 
NGC 1851 HB is considerable.

The CMD of NGC 1851 has most recently been studied by Walker (1992) (hereafter W92)
in the $B$ and $V$ bands, in the UV by Parise et al (1994) and in the $V$ and $I$
bands by Saviane et al. (1997) (hereafter S97), where references to earlier work can be 
found. In both optical studies the bimodal HB is interpreted as a consequence 
of differing efficiencies of mass loss as the stars evolve up the RGB.  W92 
suggested that a unimodal mass distribution might be able to produce a bimodal 
HB stellar distribution when the detailed shape of the evolutionary tracks was 
taken into account, however S97 do not find good agreement when comparing with 
the Bertelli et al. (1994) tracks.  They prefer a bimodal mass loss 
distribution, and indeed find some evidence to suggest that the radial 
distributions of the blue and red HB stars differ, pointing towards some 
as yet unexplained interaction between the dynamical evolution and the 
stellar evolution of these stars.  On the other hand, Catelan et al (1998), using 
updated (Sweigart 1997) Sweigart and Gross (1976, 1978) models, find they can reproduce the
NGC 1851 HB morphology with a unimodal, albeit very wide, mass distribution
having characteristics $<M_{HB}> = 0.665 M_{\sun}$ and $\sigma_M = 0.055
M_{\sun}$.   

Stetson et al. (1996) and Sarajedini et al. (1997) review the question of the relative 
ages of globular clusters, to reach very different conclusions.  In both cases the
HB level of NGC 1851 (W92) is used to link the second-parameter pair NGC 
288 and NGC 362.  Critical to these arguments is the V magnitudes of the reddest 
BHB stars and the RR Lyraes; the latter will be provided here for the first time. 
 
Lying in the region of the HB which is sparsely populated, the RR Lyraes may 
provide important clues to help explain the reason for the bimodal HB, given the 
constraints that the pulsation properties place on stellar evolutionary status.
There have been suggestions (Catelan 1997) that the 1851 variables are peculiar
with respect to their behavior in the period-temperature diagram and that the 
photographic studies, as detailed below, also show that several of the RRab stars 
have light curve amplitudes near 2 magnitudes, much larger than normal. 

Sawyer Hogg (1973) lists 10 variables in NGC 1851, from discoveries by Bailey (1924) 
and Laborde and Fourcade (1966).  Preliminary periods for some of these stars, and 
an additional four new discoveries, resulted from a short observing campaign by Liller 
(1975) using photographic photometry at the CTIO 1.0-m and 1.5-m telescopes.  She noted 
that V2 and V8 appeared to be constant, and V9 was very red.  Wehlau et al. (1978) 
(hereafter W78) measured an additional 57 plates, almost all taken with the 1.0-m Swope 
telescope at Las Campanas, and analysed them along with the Liller (1975) plates.  Periods 
were derived and light curves presented for a total of 19 RR Lyrae stars, the mean period 
of the RRab stars being 0.573 days, and the ratio of the number of RRc to RRab variables 
was found to be 0.36, both are values typical of an Oosterhoff type I system.  The 
photometric zeropoint calibration for these observations was very uncertain, due to 
the lack of a definitive photometric sequence in the field.  Stetson (1981) found four 
additional variables and two apparently constant stars lying in the instability strip.  
Wehlau et al. (1982) (hereafter W82) studied these stars using their original plate material 
supplemented with another 18 plates taken in 1970, whereupon three stars were found to be 
RR Lyraes, a fourth was classified as a probable field W UMa star, while a fifth is a red 
variable.  W78 and W82 determined accurate periods and approximate mean $<B>$ magnitudes 
for all the stars they identified as RR Lyraes, classifying 15 stars as RRab and seven as RRc.  
The light curves display the $0.1-0.2$ mag scatter typical for the photographic technique. 

No modern studies of the NGC 1851 variables appear 
in the literature, however S97 identify an additional seven candidate RR Lyrae variables 
from a comparison of their photometry with that of W92, selecting stars with deviant 
photometry and $V \sim 16, V-I \sim 0.4$.  
Here we present new CCD photometry for the NGC 1851 RR Lyrae variables and compare with 
results for RR Lyraes in other GC's in this program (IC 4499, M68, M72, NGC 6362).

\section{Observations and Data Reduction}

All observations were made using the CTIO 0.9-m telescope and Tektronix 2048 \#3 CCD, 
during five observing runs in 1993 and 1994.  NGC 1851 was approximately centered on the 
CCD and exposures taken in order $V, B, I$.  Exposure times were always 300s, 600s, and 
300s respectively, with the CCD being read out through either two or all four amplifiers 
simultaneously, using an Arcon CCD controller.  Field size was 13.6 x 13.6 arcmin with 
pixel scale 0.40 arcsec.  Further observational parameters are listed in Table 1, a total 
of 126 frames in each color were reduced.  It should be noted that although the same filter 
set was not used throughout, the components for the $B$ and $V$ filters came from the same 
melts and the resulting filters have near identical passbands. Similarly, the two $I$ 
filters are also near identical.  On each photometric night several standard fields were 
observed (Landolt 1992).

The raw CCD frames were processed by zero-level (bias) subtraction followed by flat field 
division, the flats being short exposures of the twilight or dawn sky.  Calibration frames 
were built from combinations of several individual frames with clipping of cosmic rays and 
interpolation over bad columns.  All exposure times were sufficiently long that the center 
to corner shutter timing error was $\ll 1\%$.  These procedures produced object frames with 
the sky flat to better than 0.5\% in all filters.

Stars were identified and measured using DAOPHOT and ALLSTAR (Stetson 1987, 1995).  Due to 
the large number of frames, the reduction programs were run in batch mode, via a script.  
The input to this script consisted a list of frames, the full-width at half-maximum (fwhm) 
of a typical star profile for each frame, the approximate x and y coordinates of the cluster 
center, and radii of an annulus which excluded the crowded stars near the cluster center.  
For NGC 1851 potential point-spread function (psf) stars closer than 90 arcsec from the cluster 
center were rejected.   Each frame was then processed in turn by the script.  The input fwhm 
was directly utilized in the ``find'' algorithm, and also used to generate the fitting radius 
and size of the sky annuli for ALLSTAR, and the size of the box for the psf stars.  The ``pick 
psf'' routine was modified to only accept $200-300$ candidate psf stars from the specified  
annulus.  The initial pass through the frame produced a list of stars and determined an approximate 
psf, chosen to be a Moffat function plus look-up table of residuals, permitted to vary 
linearly with position on the frame.  ALLSTAR was then run, followed by subtracting all the 
fitted stars with the exception of the psf stars.  A new psf was determined, this time allowed 
to vary cubically with position, and extra stars not found on the first pass through the data 
were added to the star list.  ALLSTAR was then re-run on this larger list, utilizing the iterated 
psf.  In conditions of good transparency, these procedures typically measured between 4000 and 
15000 stars on each frame, depending on the seeing which critically affected the number of 
stars measured near the cluster center. Away from the central regions the magnitude limit was 
typically $V \sim 21$, approximately five magnitudes below the level of the HB.  Examination 
of the star-subtracted frames output from the final pass through ALLSTAR showed good removal 
of the fitted stars, even those afflicted with severe coma near the corners of the CCD frames.  
This was particularly true for the June 1995 observations, which followed repairs to the 0.9-m 
telescope primary mirror radial supports and consequent improvement in image quality.

Matching up of all the approximately 3 million measurements was achieved using the programs 
DAOMATCH and DAOMASTER (Stetson 1995), which produced lists of stars for each frame, with 
their photometry, on a common numbering system.  At this stage color equations (Table 2) 
and zeropoint adjustments were applied for each trio of $V, B, I$ frames, via a set of 
local standards.  On each photometric night sufficient Landolt (1992) fields were observed 
to measure color equation coefficients, extinction coefficients and zeropoints.  The formal 
errors on the latter are typically a few millimags in each of $V$, $B-V$ and $V-I$, and 
calibration errors arising from the standard star photometry alone should not exceed 
0.01 mag for stars of normal colors ($B-V \sim 0-1.5$).   The success in transfering aperture 
photometry of the local standards on the NGC 1851 frames, measured in exactly the same way 
as the primary standards, can be ascertained best by comparisons of results from frames 
taken on different nights and different runs, often under conditions of very different 
seeing.  The local standards have already been chosen to be relatively bright and as 
uncrowded as possible, so there should not be systematic differences between aperture and 
psf photometry due to faint stars in the star aperture.  In both aperture and psf photometry 
faint stars in the sky background have little effect for these bright stars, avoiding 
most of the problems involved with sky determination (eg Stetson 1987).  Notwithstanding, 
linking the local standard magnitudes to the primary standards is generally acknowledged 
to be the step most likely to introduce systematic calibration errors, and is perhaps 
best checked by comparing with completely independent sets of photometry. 
   
There is extensive accurate photometry for stars in the region of NGC 1851.  For $V$ and 
$B-V$, W92 found from inter-comparsions between his own photometry of five bright 
stars near NGC 1851 with that by Stetson (1981), Alcaino et al. (1987), Sagar et al (1988), 
and Da Costa and Armandroff (1990) ($V$ only) that agreement for $V$ and $B-V$ is no worse 
than $0.01 - 0.02$ mag for all these measurements, and in particular the zeropoint offset 
between the W92 and the Stetson (1981) photometry is $0.00 \pm 0.01$ mag for both 
$V$ and $B-V$.  On the basis of this excellent agreement he chose a set of $15-18$ mag stars 
as local standards.  Here, due to the smaller telescope aperture, we prefer to use brighter 
local standards ($V = 13-15$) although there are some stars in common.  Apart from the present 
photometry, we have $V, V-I$ photometry by S97, Da Costa and Armandroff (1990) 
and Alcaino et al (1987, 1990, hereafter A90).

We see no reason to alter the $V$ and $B-V$ zeropoints as used by W92, as the new calibrations 
reproduce the W92  $V$ and $B-V$ zeropoints to within $\pm 0.01$ mag.  The situation with 
respect to the $V-I$ photometry is more complicated.  The $V-I$ calibrations from the present 
data set reproduce both the bright A90 stars and the fiducial RGB of Da Costa and Armandroff 
(1990) to within $\pm 0.01$ mag.  The A90 fiducial for the lower RGB, subgiant branch and MS 
lies a mean of 0.02 mag redder than the present sequence to as faint as $V = 20.5$.  S97 state 
that their $V$ magnitude zeropoint agrees with W92, nominally to $0.001 \pm 0.001$ mag (internal 
error), and with similar good agreement (0.012 mag) with A90.  However when they compare their 
$V-I$ zeropoint with A90 they find an offset of $0.035 \pm 0.004$ mag in the sense that S97 are 
redder.  Now S97 state that their external calibration errors are 0.03 mag in $V$ and 0.02 mag 
in $I$, so the $V$ agreement may be fortuitous and the $V-I$ zeropoint offset is only slightly 
larger than a one sigma error.  The present $V-I$ measurements when compared to S97 actually 
show good agreement (to 0.01 mag or better) for the HB and on the RGB to the level of the HB.  
Fainter than this there are systematic offsets between S97 and both A90 and the present work, 
which grow larger at fainter magnitudes in the sense that S97 is redder.  On the MS at $V = 20$, 
S97 is 0.07 mag redder than A90 and 0.08 mag redder than here.   Now S97 has significantly better 
image quality (seeing 1.1 - 1.2 arcsec) for their frames, albeit with rather short exposure times, 
and so the source of the non-linearity is not certain.  An indication might be that somewhat 
similar behavior is seen when comparing the B-V colors of the MS here with those of W92, where 
W92 is redder by 0.02-0.03 mag despite excellent agreement for brighter stars.   In any case, the 
systematic trends at faint magnitudes are irrelevant for the present work,
and in fact it is gratifying that the agreement for HB and brighter stars is so good.  NGC 1851 
is a cluster with high central concentration, and ground-based photometry within an arc minute 
of its center is difficult, even with excellent seeing.  

To summarize, we believe that the $V$, $B-V$, and $V-I$ magnitude zeropoints are known to better 
than $\pm 0.02$ mag, and probably to $\pm 0.01$ mag.  The various sources of CCD photometry agree 
well with each other to $V \sim 18.5$, but fainter than that some systematic trends with magnitude 
occur.

\section{RR Lyrae Photometry}

Candidate variable stars were identified using the Welch \& Stetson (1993) method, which 
looks for correlated changes in near-simultaneous measurements made at two wavelengths.  
Since we have observations made near-simultaneously at three wavelengths, the {\it variability 
index} was rewritten as

\[I \propto \sum_{i} \delta B  \delta V  \delta I \]

where the $\delta$ magnitudes are the differences with respect to the mean.  In order to 
strengthen discrimination against ``bad'' measurements biasing the variability index,
recognition was taken of the fact that for RR Lyraes the amplitude ratios in the three filters are
on average $B:V:R = 1.0:0.8:0.5$ thus the delta magnitudes were required to obey this condition, 
with generous limits, otherwise the observation was rejected.  Thus not only must variations at 
a given time be correlated between the three colors, but the ratios of the variations in each 
band must be reasonable.

At the same time, mean magnitudes were calculated for all the stars.  Weighting
the individual results via the photometric errors returned by ALLSTAR is not a
useful technique for calculating mean magnitudes in such a crowded field, where
the major source of scattter is incorrect measurements of blended stars, particularly 
on the poorer-seeing frames.  Instead, residuals about an initial
mean were calculated, and the most deviant values discarded.  This process was
iterated, until convergence was achieved with more than half the measures
remaining, otherwise the star was discarded.  Table 3 includes all stars measured brighter than 
magnitude $V = 19$ and contained within an annulus of radii 80 and 260 arcsec centered on 
the cluster.

The program produced a list of variable stars, prioritized by the {\it variability index}.  
The stars were then subject to period-finding using a phase-dispersion minimization program 
and a least-squares fitting periodogram program, both written originally by Dr L A Balona (SAAO).
As a result of these procedures all 22 previously known RR Lyrae variables were re-discovered, 
along with another seven additional stars, six RRab and one RRc.   The mean period of the RRab
stars increases slightly, to 0.586 days, and the ratio RRc:RRab is now 0.38.  The new discoveries
are numbered V27 through V33.  We confirm (W78, W82) that V2, V9 and V24 are 
red stars, and that V25 is a likely field W UMa variable.  The full CCD field (13.6 arcmin square) 
is shown in Figure 1, with the more distant RR Lyraes identified.  Figure 2 shows a 100 arcsec 
square field centered on the cluster, with the inner RR Lyraes marked.  Photometry
within 30 arcsec of the cluster center has large systematic errors and a CMD for these 
stars shows much scatter.  Notable is star no. 17, only 25 arcsec from the cluster center, 
and which can be identified on Figure 2 as the bright star 7 arcsec N and 1.6 arcsec W of V31.  
This star has $V = 13.21, B-V = -0.07, V-I = -0.12$ and is UV-5 (Vidal \& Freeman 1975) a radial 
velocity member and thus a star in the post-asymptotic giant branch (PAGB) stage of evolution, 
(see S97 figure 6).  S97 identified several ``supra-HB'' candidates but most of these are well-distant 
from the cluster, and will require membership confirmation.  No bright stars with $B-V < 0.6$ are here 
found within 4 arcmin of the cluster center, and indeed two other PAGB candidates (UV-6, UV-7) have 
already been shown to be radial velocity non-members (Da Costa 1982).

The high concentration of NGC 1851, coupled with the fact that only 206 stars, mostly brighter 
than the HB, were able to be measured within 25 arcsec of the cluster center, means that it 
is certain that more RR Lyraes exist in the very crowded central regions.  Their discovery 
will require much better image quality that the $1.3-1.8$ arcsec fwhm typical for the stars 
on the frames available here.  Of the seven stars suggested by S97 as candidate variables, 
three are amongst the seven new variables mentioned above, while the other four stars are 
constant.  Two of the latter are close ($\sim 1$ arcsec) doubles.

The RR Lyrae $V$ magnitude light curves are plotted as a function of phase in Figure 3.  
In almost all cases period-finding was straight-forward, however several stars show poor-quality 
light curves, particularly those nearer to the cluster center.  In particular, V30 shows 
point-point scatter of more than 0.1 mag and the two longest period RRc stars, V19 and V33, 
are scarcely better.  The shortest period RRc star, V23, appears to have variable amplitude, 
while the RRab stars V5, V29 and V28 probably display the Blazkho effect.  V10 proved to be 
particularly troublesome.  It has a period by W78 of 0.49948 days but this does not fit the 
present data set, and 0.49975 days is preferred.  This fits the data better, but not perfectly, 
and several other candidate periods were investigated, with no success.  Aliasing is a problem, 
and a period near 0.333 days is also possible.  The rather low amplitude of the light curve 
would support this choice, but the color of the star and the better fit of the chosen period 
would appear to rule out this possibility.  A long period RRab option, which would be 
consistent with both colors and amplitude, does not seem possible from examination of the 
power spectrum.  Three stars have photometry severely contaminated by close companions, 
these are V14 which is brighter than expected, and has a clearly elliptical profile, 
while star V30 is very crowded, being only 15 arcsec from the cluster center.  
Star V19 is also rather bright, with a fainter close companion, and photometry 
for all these stars should be treated with caution.  Individual measurements
for the RR Lyraes are given in Table 4, which lists the heliocentric Julian date corresponding 
to the midpoint of the $V$ exposure, the corresponding phase with arbitrary zero, the additive 
offsets in days to the midpoints of the B and I exposures, the $V, B, I$ magnitudes, and the 
errors in the $V, B$ and $I$ magnitudes.  

The data set are not optimal for finding other than short period variables, although stars 
varying on timescales of months can be noticed when observations from the different observing 
runs are compared.  Five such stars at the tip of the RGB were found to be variable, with 
pseudo-periods of timescale a year and amplitudes of $0.1-0.2$ mag.  No variable blue 
stragglers were found.  

\section{RR Lyrae Derived Quantities}

The technique of deriving astrophysical quantities from Fourier decomposition of the light 
curves of RR Lyrae stars was pioneered by Simon (1988).  Simon and Clement (1993) compared 
their hydrodynamic models with Fourier parameters for RRc stars and derived expressions 
evaluating mass, luminosity, temperature and relative helium abundance.  Jurcsik and Kov\'{a}cs 
(1996) and Kov\'{a}cs and Jurcsik (1996) found relationships between the Fourier parameters of 
RRab stars and their metallicity and luminosity.  These results have recently been tested 
against observations of RR Lyraes in seven galactic GC's by Clement and Shelton (1997).  
The derivation of luminosities from the Fourier coefficients for the NGC 1851 and other GC RR 
Lyraes will be discussed in detail elsewhere (Kov\'{a}cs and Walker, work in progress). 
Fourier series were fitted to the light curves, to evaluate intensity- and
magnitude-mean magnitudes, mean colors, amplitudes, as well as the Fourier amplitude and
phase coefficients, as described by Walker (1994).   We have calculated coefficients for 
all three ($V, B, I$) lightcurves, which allow assessment of the coefficient errors 
independent of the formal errors returned by the fitting program, which are also tabulated.  
Intensity mean magnitudes, mean colors, and amplitudes are listed in Table 5, while Table 6 
contains the magnitude means, the Fourier phase coefficients and combinations.  The very 
large amplitudes found for some stars (eg V1, V8, V16) by W78 are not confirmed.  We note 
that their photographic photometry is based on a photoelectric sequence by Alcaino (1971) 
that has some uncertainties at the faint end, which is itself a magnitude brighter than the 
RR Lyraes at minimum light.  For stars brighter than $B\sim 16$, the photographic photometry 
shows good agreement with the CCD work, but is fainter by 0.5 mag by $B \sim 17$. This is 
mostly likely due to an incorrect calibration of the photographic plates when extrapolating 
from the fainter photoelectric sequence stars.

Bono et al. (1995) have calculated corrections to observed mean colors to obtain the color 
of the equivalent static atmosphere, as a function of light-curve amplitude.  For the 
static mean $V$ magnitude, the corrections to be applied to the intensity mean magnitudes 
$<V>$ in Table 5 are very small, even for the highest amplitude RRab star, V1, the correction 
is only -0.01 mag.  It is worthwhile to note that the correction required to the magnitude 
mean $V$ for this star is -0.08 mag.  If these corrections are applied, then in both cases 
the equivalent static $V$ magnitude is 16.04, a gratifying result.   The corrections required 
for $(B-V)_{mag}$ are also not large. RRab stars with amplitudes $A_B$ in the range $0.7-1.4$ 
mag  need a correction of -0.01 mag, rising to -0.02 mag by $A_B = 1.7$ mag.  For the very low 
amplitude RRc star the correction is +0.008 mag, for all other RRc stars the correction is 
within a few millimags of -0.005 mag.

The question of how to convert between RR Lyrae colors and effective temperatures $T_e$ is a vexing one.  
The first step, that of calculating the correct mean color from those observed, has been dealt 
with in the preceeding paragraph.  Kurucz (1992) calculated colors as a function of $[Fe/H]$, $T_e$, and
log g from model atmospheres, and these tables have been widely used.  For RR Lyraes, Sandage (1981) 
first suggested that the blue amplitude $A_B$ measured relative temperatures for RRab stars, and this
approach was extended by Carney et al. (1992) who found a relationship $T_e = f(A_B, [Fe/H], P)$ best 
represented temperatures for their sample of RRab stars.  Catelan et al. (1998) argue that 
a relationship independent of P is preferable, and show that their temperature scale 
is in excellent agreement with the Carney et al. (1992) temperatures.  However the use of this relation
in period-shift analyses shows much more scatter than if Carney et al. (1992) temperatures are used.
Castellani and de Santis (1994) note that the metallicity dependence is 
slight, and prefer to define their temperature scale in terms of P and $A_B$ alone. 
In general, these relations are not very robust given the number of parameters
compared to the calibrating stars.  It thus seems preferable to use the colors directly, and recently
Castelli, Kurucz, and Gratton (1997a,b) (hereafter C97) 
have provided calibrations based on revised model atmospheres.
McNamara (1997) has carefully compared consistency between the temperatures so predicted from various 
colors, and the consequently derived gravities and luminosities.  He concludes that the optical
colors best correspond to ``correct'' effective temperatures, and that the oft-used V-K scale is 
systematically offset.  The optical temperature scale is some 200-300 K hotter for a given color than 
the older Kurucz (1992) scale, and he goes on to show that the higher temperatures and consequent
higher luminosities are consistent with a distance modulus of 18.53 mag for the Large Magellanic Cloud,
similar to that given by other distance indictators.  We will use the C97 temperature
scale here, interpolating in the provided table to $[Fe/H] = -1.29$ and log $g = 2.75$ (Fernley 1993),
relevant for the NGC 1851 RR Lyraes.  The B-V colors will be used to best allow comparisons with 
earlier work.

\section {Reddening and Metallicity}

It is convenient to deal with reddening and metallicity together, since most photometric methods do 
not determine these two quantities independently.   W92, after an extensive discussion of the 
available reddening measurements, concluded that there was very little evidence to support the 
sometimes rather high values suggested by some workers, and adopted $E(B-V) = 0.02 \pm 0.02.$   
In similar manner, W92 found $[Fe/H] = -1.29 \pm 0.07$, as advocated by Da Costa and Armandroff 
(1990), was consistent with the new photometry.  S97 adopted the same value, which is on the 
Zinn and West (1984) scale, used here for consistency with previous work.  Rutledge et al. (1997), 
in their discussion of GC $[Fe/H]$ derived from measurements of the \ion{Ca}{2} triplet lines, find 
that the recent Carretta and Gratton (1997) metallicity scale gives $[Fe/H]$ values some 0.2 dex 
more metal rich than the Zinn and West (1984) scale for clusters with metallicity similar to NGC 
1851.  They also conclude that it is not at all obvious which of the two scales best measures 
$[Fe/H]$.  

As noted earlier by Armandroff and Da Costa (1990), the Ca triplet measurement for NGC 1851 
lies 0.15 dex off the calibration defined by other GC if $[Fe/H] = -1.29$ is assumed, by 
approximately 0.15 dex, in the sense that the Ca triplet measurements suggest a more metal 
rich value.  Whether this represents enhancement of $[\alpha/Fe]$ relative to the other 
GC that calibrate the Ca triplet metallicity scale, or incorrect observations, would be 
important to resolve with a new spectroscopic determination.

We can measure the reddening and metallicity here from the position of the RGB in the $V, 
V-I$ CMD (Sarajedini 1994), or in the $V, B-V$ CMD (Sarajedini and Layden 1997), and also 
calculate the reddening via Sturch's method (Sturch 1966, Walker 1990, Blanco 1992) applied 
to the RRab variables.

We use Sturch's method as described by Walker (1990), where the reddening zeropoint has 
been adjusted to correspond to zero reddening at the galactic poles and the [Fe/H] scale 
is that of Zinn and West (1984), whereby

$E(B-V) = (B-V)_{min} -0.24 P -0.056 [Fe/H] - 0.336$,

where $(B-V)_{min}$ refers to phases $0.5 - 0.8$ and is applicable for RRab stars only.   
For 13 stars with $\sigma (B-V)_{min} < 0.04$ mag, and with $[Fe/H] = -1.29$,

$E(B-V) = 0.05 \pm 0.02$.

This is 0.03 mag higher than that determined by W92.  It should be noted however that in 
two other clusters studied in this series, M68 (Walker 1994) and IC 4499 (Walker and Nemec 
1996), $E(B-V)$ found by Sturch's method was in both cases 0.02 mag larger than the final 
adopted mean from several methods.  An alternative calibration by Blanco (1992) also produces 
reddenings smaller by about 0.015 mag.  The present result, although suggesting a slightly 
higher reddening for NGC 1851 than found by Walker (1992), should not be afforded excessively 
high weight due to these zeropoint uncertainties.
 
Sarajedini (1994) introduced a method to determine $E(V-I)$ and $[Fe/H]$ simultaneously, 
from the position and shape of the RGB in the $V, V-I$ CMD.  The method is calibrated from 
six GC, one of which happens to be NGC 1851, from photometry by Da Costa and Armandroff 
(1990).  Given that our photometry of the NGC 1851 RGB reproduces that found by Da Costa 
and Armandroff (1990), applying their method just forces NGC 1851 to fall exactly on 
the Sarajedini (1994) calibration fiducials, with $[Fe/H] = -1.4 \pm 0.2$ and $E(B-V) = 
0.02 \pm 0.02$, where $E(V-I) = 1.25 E(B-V)$.  Similarly, Sarajedini and Layden (1997) 
use NGC 1851 data from W92 as one of their primary calibrators in the $V, B-V$ version 
of the method.  In this case the position of NGC 1851 relative to the fiducials would 
argue for little change in the nominal $[Fe/H]$ and $E(B-V)$ values.

Finally, we can calculate $[Fe/H]$ from the Fourier decomposition of the RRab light curves.  
The cluster metallicity can be derived from the Fourier coefficients of the RR Lyraes (Jurcsik 
and Kov\'{a}cs 1996). who derive a linear relation between $[Fe/H]$, Period $P$ and the Fourier 
phase parameter (see Simon 1988) $\phi_{31}$,

 $[Fe/H] = -5.038 - 5.394P + 1.345 \phi_{31} .........................(1)$

where because of the $2\pi$ ambiguity the phase should be taken closest to the sample 
average, 5.1.  Note that Simon (1988) decomposes the Fourier series as cosines (as are 
the values in Table 6) whereas Jurcsik and Kov\'{a}ks (1996) prefer a sine decomposition.  
To convert from cosine to sine decomposition requires subtracting $\pi / 2$, $3\pi /2$, 
and $5\pi /2 $ respectively from the phase
combinations $\phi_{21}$, $\phi_{31}$ and $\phi_{41}$.  We proceed by choosing
the nine RRab stars with the best light curves, and weight $\phi_{31}$ found from 
each of $B$ and $V$ twice that of the $I$ result, since the lower amplitude $I$ light 
curves always give a much larger error for $\phi_{31}$ than do B and V.  It is found that

$[Fe/H] = -1.31 \pm 0.05$ (s.e.).

In this case the $[Fe/H]$ scale is that described by Jurcsik (1995), which is based on 
high dispersion spectroscopy rather than the Zinn and West (1984) scale.  An approximate 
conversion to the Zinn and West (1984) scale follows from noting that M4 has $[Fe/H] = -1.28$ 
on the latter scale, while four independent high dispersion spectroscopy measurements give 
mean $[Fe/H] = -1.11$.  This suggests that the metallicity on the Zinn and West (1984) scale 
for NGC 1851 is near $[Fe/H] = -1.45$ via equation (1).

In summary, the new estimates for the metallicity and reddening of NGC 1851 are close to 
the values of $[Fe/H] = -1.29 \pm 0.07$ and $E(B-V) = 0.02$ adopted by W92 and S97, and 
these values will be retained for this paper.  The question of the {\it true} GC $[Fe/H]$ 
scale is obviously a very important one, and given that the Zinn and West (1984) 
and Carretta and Gratton (1997) scales differ non-linearly, ramifications when comparing 
stellar properties between clusters of differing metallicities will be pervasive (Jucsik 1995).

\section{Period Shift}

We exclude V14 from this analysis as its magnitude is clearly discrepant, see
above. Temperatures 
were calculated from C97 models, as described above, then periods 
for the first overtone pulsators were adjusted to the equivalent fundamental period by
the addition of 0.125 to log $P$ (van Albaba \& Baker 1971).
A further correction, to compensate for the period
change due to evolution in luminosity, is usual.  Brocato et al. (1996) argue that the
ZAHB should be the reference luminosity for this correction, rather than the mean magnitude
(Sandage 1990).  The latter has the disadvantage of being sample dependent, although simple
to calculate, whereas care is needed to observationally define the ZAHB position correctly.
With a reasonable sample of RR Lyraes, as is the case here, the ZAHB can be accurately 
defined.  A mean of the faintest 11 stars is $V = 16.115 \pm 0.002$ (internal error), and a
true lower bound would be only 0.01 mag fainter.  For comparison, we will use the photometry
and periods determined for M68 (Walker 1994), IC 4499 (Walker \& Nemec 1996) and NGC 6362
(in preparation), calculating periods and temperatures in the same way as described for 
NGC 1851.  For M68, $V_{ZAHB} = 15.702 \pm 0.005$, for IC 4499 $V_{ZAHB} = 18.29 \pm 0.01$,
and for NGC 6362 $V_{ZAHB} = 15.33 \pm 0.01$.  Period-shift diagrams are plotted in Figure 
4.  The observational uncertainty mostly affects the temperature axis, as in all these clusters
the ZAHB can be well-defined, and an (unlikely) error of placement of the ZAHB of 0.03 mag
corresponds to only a displacement of 0.01 in log P.  Once the ZAHB is chosen, magnitude
measurements are differential.  The major source of observational error
occurs in the calculation of the temperature, where an error of 0.01 in log T corresponds
to 0.03 mag in the B-V color, which must first be corrected for reddening.  The figure shows no
period shift between IC 4499 and NGC 1851, and only an insignificant indication of a shift
between these two clusters and NGC 6362.  Since NGC 6362 and IC 4499 bracket NGC 1851 in
metallicity, we conclude from this diagram that the period-shift behaviour of the NGC 1851
variables appears normal.  On the contrary, the M68 variables have a substantial period-shift
when temperatures are calculated using E(B-V) = 0.07, as was found by Walker (1994) when
comparing to the OoI cluster M3 using a different 
color-temperature  calibration (Sandage 1990).  The shift could be completely removed if the 
M68 reddening was reduced to $E(B-V) = 0.03$, the value favoured in the Brocato et al. (1994) 
study of M68 (see Figure 4).  Walker (1994) provided several reasons why the higher value is 
to be prefered, but re-evaluations by Gratton  et al. (1997)and especially by Brocato et al. (1997) would argue for a value of
$E(B-V) = 0.04 \pm 0.01$.  Gratton et al. (1997) also prefer a slightly higher metallicity for
M68, $[Fe/H] = -2.0$, but the 0.1 dex change has an insignificant effect on the temperatures.
Final resolution of the M68 reddening question would appear to require further observational effort.

\section{CMD morphology}

The CMD in general has been described by W92 and S97, the latter also include a
pre-COSTAR HST CMD of the cluster center.  A more recent HST CMD of the 
central region of NGC 1851 is presented without discussion by Sosin
et al (1997).   These CMD's are all deep enough to show stars well below the MS 
turn-off.   
In Figure 5 we plot all stars measured brighter than $V \sim 20.5$ in an annulus 
with radii 80 and 260 arcsec, while Figure 6 adds all measured stars to within 30 
arcsec of the cluster center.  This includes all the known RR Lyraes except V30 
which is only 15 arcsec from the cluster center. Both diagrams show only a few
field stars well-distanced from the cluster principal sequences, so field star
contamination is not an issue when interpreting these diagrams.  As expected,
Figure 5 shows fewer blends scattered off the principal sequences.

The RGB is well-populated and clearly separated from the asymptotic giant branch (AGB).  
At fainter magnitudes, and at bluer colors than the MS, are a number
of blue stragglers, which are extensively discussed by S97.  The MS itself is well-defined, 
by considering the Figure 5 stars with $19.6 < V < 19.1$, then $\sigma(B-V) = 0.015$ mag 
for a best-fit Gaussian, only slightly larger than the
mean internal photometric error for these stars of 0.009 mag.   The true external scatter 
due to photometric errors will be larger than this figure by some not well-determined amount, 
even though the increase in color dispersion arising from blends and binaries is minimized by 
chosing the near-vertical part of the MS where blending MS stars all have very similar colors 
and thus contibute little to the color dispersion.   Adding 0.005 mag in quadrature to the 
internal errors restricts the possible intrinsic width of the MS to the equivalent 
of a range in metallicity of approximately 0.1 dex.  A larger and probably more realistic 
internal to external error correction effectively allows for no dispersion in metallicity, 
which is also the conclusion reached by S97.

The HB shows a
very condensed clump of red HB stars, with a pronounced gap on the red side. Several
stars are loosely grouped above and slightly to the red of the red HB, these stars are 
presumably in a state of evolution at the very end of, or beyond, core helium burning.
One or two of these stars may be evolved blue stragglers.  
The star density falls steadily bluewards along the HB, reaching a minimum in the 
vicinity of the RRc stars, before increasing again with an extensive tail of  blue 
HB stars extending to $V \sim 17, B-V = -0.05$.  A well-measured, very blue star 
at  $V = 18.46, B-V = -0.132$ is only 120 arscec from the cluster center and
thus has a good probability of being a member.  This star (UIT 31) is one 
of two with very blue UV colors (Parise et al 1994), indicative of temperatures
well in excess of 10000 K.  The second star (UIT 44) has even bluer UV color than
UIT 31, but Landsman (1994) found rather unremarkable optical colors ($V =
18.6, B-V = 0.11$) and confirmed the star as being a radial velocity member 
of NGC 1851.  He suggested that perhaps the star was a binary, and indeed a 
combination of a star near the MS turn-off ($V = 19.2, B-V = 0.6$) when subtracted
from the optical colors gives $V = 19.5, B-V = -0.3$, much more consistent with
the UV colors and the optical spectra, both of which indicate a 
temperature near 30000 K.

The ZAHB is very cleanly defined, and is not {\it horizontal}, being slightly brighter at 
bluer colors.   This is best seen in Figures 7 and 8, where the ZAHB appears to
slope steadily upwards towards the blue, from V = 16.20 at B-V = 0.65 to
V=16.12 at B-V = 0.20.  Bluer HB stars become rapidly fainter for a small
change in color, as B-V becomes insensitive to temperature changes and the
bulk of the emitted radiation from these hot stars occurs in the UV.  In Figure 7,
four stars lie slightly above the blue HB, near $V = 16.0, B-V = 0.08$.  Three
of these stars are crowded.  The fourth, which lies 111 arcsec from the cluster
center and ought to be relatively uncrowded, actually has a red star almost two 
magnitudes brighter in $V$ at 3 arcsec south, and a slightly fainter star at separation
1.5 arcsec west, so the photometry may well be incorrect.  Apart from these stars,
and the few RR Lyraes with unreliable photometry as discussed above, the vertical
extent of the HB appears to narrow with bluer $B-V$ colors.  Catalan et al (1998) 
state that observations such as these constrain the theoretical paths of HB evolutionary tracks; 
they note that the Lee and Demarque (1990) tracks have longer blue loops for a given 
metallicity than do either their own tracks or indeed the later Yale tracks (Yi, Lee 
and Demarque 1995).  Both of the latter agree much better with the observed HB star 
distribution in NGC 1851.

No significant gaps are apparent in the star distribution along the HB. 
The numbers of blue HB stars (B), compared to variables (V) and red HB stars (R) 
are: $B:V:R = 31:15:64$ in a 30-80 arcsec annulus and $B:V:R = 39:13:61$ in an
80-260 arcsec annulus.  Assigning errors of $\pm 2$ counts for the non-variables, and 
$\pm 1$ counts for the variables, then we can compare relative numbers.  For the whole 
sample, the Lee parameter $(B-R)/(B+V+R) = -0.25 \pm 0.03$.  The ratio B:R in the 
inner zone is $31:64 = 0.48 \pm 0.05$, and in the outer zone $39:61 = 0.64 \pm 0.05$.  
The difference is barely significant.  A more stringent investigation by S97 found 
that 2-population Kolmogorov-Smirnov tests indicated that within 100 arcsec from the 
cluster center the RHB, BHB and SGB populations all had the same 
distribution but outside 100 arcsec the BHB stars appeared slightly less concentrated.
We have also counted RGB stars, down to V = 16.0 to avoid the clump stars.  Generally, 
the AGB stars are easy to identify, and field contamination is negligible (Ratnutunga
and Bahcall (1985).   We count 60 RGB stars in the 
30-80 arcsec annulus, and also 60 stars in the 80-260 arcsec annulus, with an error 
of $\pm 4$ stars each.  We have included the very red star 59 ($V=13.39, B-V = 1.88, 
V-I = 1.89$) as a member, the next reddest star is 63 ($V = 13.40, B-V = 1.58, V-I = 1.58$).   
The ratio of RGB stars in the two annuli is not significantly different from the 
corresponding ratios for the RHB and the BHB stars, although nominally closer to 
that for the RHB stars.

The edges of the instability strip, judged by the measured colors
of variables near the strip boundaries, are $(B-V)_0 = 0.19 \pm 0.01, 
(V-I)_0 = 0.235 \pm 0.01$  corresponding to the first overtone blue edge, and 
$(B-V)_0 = 0.41 \pm 0.01, (V-I)_0 = 0.545 \pm 0.01$, corresponding to the 
fundamental red edge.  The  RRc/RRab boundary is at $(B-V)_0 = 0.29 
\pm 0.02, (V-I)_0= 0.365 \pm 0.02$.  These will first be compared with the results 
for some other clusters, and also with theory.   

It has long been suspected (eg Deupree 1977) that the red edge of the instability 
strip moves to redder colors at higher metallicities. This has been difficult to
confirm as the postulated shift is not large, comparable to the typical errors in
reddening determinations.  Instability strip boundary colors for several galactic and 
LMC clusters observed in similar manner to NGC 1851 are listed in Table 7.  A least
squares fit to the the colors of the red edge shows a shift of $0.04 \pm 0.02$ mag/dex,
over a metallicity range of approximately $[Fe/H] = -1$ to -2.  This is only a
$2 \sigma$ result and data from more clusters are clearly needed.
The corresponding temperatures
can be derived from the C97 tables, assuming log $g=2.75$ 
(Fernley 1993) and 
show that the temperature change is not large, falling from 6230 K to 6120 K.  For reference,
at fixed metallicity and gravity this temperature change would correspond to only 0.03
mag.  By contrast, the color of the blue edge appears constant near $B-V = 0.18$,
however the corresponding temperatures are 7300 K at $[Fe/H] = -2$ and 7360 K at $[Fe/H] 
= -1$.  

We can compare these results with theoretical predictions by Bono et al. (1997),       
who have calculated mode boundaries for $Z=0.02$ and $Z=0.0001$ models. The slope 
of their bolometric 
magnitude-metallicity relation is 0.24.  Their metal-rich model has $M = 0.53M_{\sun}$
and $Y = 0.28$, the metal-poor model has  $M = 0.65M_{\sun}$ and $Y = 0.24$.  The
lack of a model with $Z=0.001$ immediately makes what follows preliminary, given the
large interpolation required.  It is immediately obvious
that there is a moderate difference in zeropoint between the predictions and transformed
measurements, since for the red edge the Bono et al. (1997) temperatures are some 300 K too
low.  The predicted temperature difference of 50 K is not too inconsistent with the 110 K 
measured, since at the red edge 0.01 mag corresponds to 35 K and there is at least an
uncertainty at this level observationally.
      
At the blue edge of the instability strip, the Bono et al. (1997) predictions are 
approximately 100K too cool.  Via the C97 transformations, a color of
$B-V = 0.18$ will correspond to a star with temperature 7360 K at $[Fe/H] = -1$ and 
7300 K at $[Fe/H] = -2$, the reverse sign to the Bono et al. (1997) prediction.  Here
0.01 mag in $B-V$ corresponds to 50 K, so again these results are not too discordant.

Finally, we investigate the color of the RRcd - RRab transition for clusters with 
sufficient stars of both groups that the position is well-defined.  This color is listed
in the final column of Table 7.  The small trend visible, that the color of the
transition reddens slightly at higher metallicities, runs counter to the trend that the
temperature (C97) at constant color increases with metallicity, thus
the transition occurs at near constant temperature, 6725 K, assuming log $g = 2.75$.

Acknowledgements:  I would like to thank Marcio Catelan for stressing the importance of the
NGC 1851 RR Lyraes, and for pointing out the unusually large photographic amplitudes.

\references

Alcaino, G. 1971, \aap, 15, 360

Alcaino, G., Liller, W., \& Alvarado, G. 1987, \aj, 93, 1464

Alcaino, G., Liller, W., Alvarado, F., Wenderoth, E. 1990, \aj, 99, 817  (A90)

Bailey, S. 1924, Harvard Bull. no. 802

Bertelli, G., Bressan, A., Chiosi, C., Fagotto, F., \& Nasi, E. 1994, \aap S, 106, 275

Blanco, V.M. 1992, \aj, 104, 734

Bono, G., Caputo, F., \& Stellingwerf, R.F. 1995, \apjs, 99, 263

Bono, G., Caputo, F., Cassisi, S., Incerpi, R., \& Marconi, M. 1997, \apj, 483, 811

Brocato, E., Castellani, V., \& Ripepi, V. 1994, \aj, 107, 622

Brocato, E., Castellani, V., \& Ripepi, V. 1996, \aap, 111, 809

Brocato, E., Castellani, V., \& Piersimoni, A. 1997, \apj, 491, 789

Carney, B.W., Storm, J., \& Jones, R.V. 1992, \apj, 386, 663

Carretta, E., \& Gratton, R.G. 1997, \aap S, 121 95

Castellani, V., \& de Santis, R.  1994, \apj, 430, 624

Castelli, F., Kurucz, R., \&  Gratton, R. 1997a, \aap, 318, 841 (C97)

Castelli, F., Kurucz, R., \& Gratton, R. 1997b, \aap, 324, 432 (C97)

Catelan, M. 1997, private communication

Catelan, M., Borissova, J., Sweigart, A.V., \& Spassova, N. 1998, \apj, in press, (astro-ph/9708174)

Clement, C.M., \& Shelton, I. 1997, \aj, 113, 1711

Da Costa, G. 1982, \pasp, 94, 769

Da Costa, G.S., \& Armandroff, T.E. 1990, \aj, 100, 162

Deupree, R.G. 1977, \apj, 214, 502

Fernley, J.A. 1993, \aap, 268, 591

Gratton, R.G., Fusi Pecci, F., Carretta, E., Clementini, G., Corsi, C.E., Lattanzi,
 M.G. 1997, \apj, 491, 749 

Jurcsik, J. 1995, Acta. Astr., 45, 653

Jurcsik, J., \& Kov\'{a}cs, G. 1996, \aap, 312, 111

Kov\'{a}cs, G., \& Jurcsik, J. 1996, \apj, 466, L17

Kraft, R.P. 1994, \pasp, 106, 553

Kurucz, R.L. 1992, in IAU Symp. 149, The Stellar Populations of Galaxies, ed. B. Barbuy \&
A. Renzini (Dordrecht:Kluwer), 225  

Laborde, J.R., \& Fourcade, C.R., 1966, Cordoba Repr., p. 138

Landolt, A.U. 1992, \aj, 104, 340

Landsman, W.B. 1994, in Hot Stars in the Galactic halo, ed. S.J. Adelman, A.R. Upgren, 
\& C.J. Adelman (Cambridge: Cambridge University press), p. 156

Lee, Y.-W., \& Demarque, P. 1990, \apjs, 73 709

Liller, M.H. 1975, \apj, 201, L125

McNamara, D.H. 1997, \pasp, 109, 857

Parise, R.A., et al. 1994, \apj, 423, 305

Ratnutunga, K., \& Bahcall, J.N. 1985, \apjs, 59, 63

Rich, R.M., et al. 1997, \apj, 484, 25

Rutledge, G.A., Hesser, J.E., \& Stetson, P.B. 1997, \pasp, 109, 907

Sagar, R., Cannon, R.D., \& Hawkins, M.R.S. 1988, \mnras, 232, 131

Sandage, A. 1981, \apj, 248, 161

Sandage, A. 1990, \apj, 350, 631

Sarajedini, A. 1994, \aj, 107, 618

Sarajedini, A., Chaboyer, B., \& Demarque, P. 1997, PASP, 109, 1321

Sarajedini, A., \& Layden, A. 1997, \aj, 113, 264

Saviane, I., Piotto, G., Fagotto, F., Zaggia, S., Capaccioli, M., \& Aparicio,
 A. 1997, \aap , in press  (S97)

Sawyer-Hogg, H.B. 1973, Publ. David Dunlap Obs., 3, no. 6

Simon, N.R. 1988, \apj, 328, 747 

Simon, N.R., \& Clement, C.M. 1993, \apj, 410, 526

Sosin, C., et al. 1997, \apj, 480, 35

Stetson, P.B. 1981, \aj, 86, 687

Stetson, P.B.,1987, \pasp , 99, 191

Stetson, P.B., 1995, DAOPHOT II User's Manual

Stetson, P.B., VandenBerg, D.A., \& Bolte, M. 1996, \pasp , 108, 560

Sturch, C.R. 1966, \apj, 143, 774

Sweigart, A.V. 1997, \apj, 474, L23

Sweigart, A.V., \& Catelan, M. 1997, preprint

Sweigart, A.V., \& Gross, P.G. 1976, \apjs, 32, 367

Sweigart, A.V., \& Gross, P.G. 1978, \apjs, 36, 405

van Albaba, T.S., \& Baker, N. 1971, \apj, 169 311

Vidal, N.V., \& Freeman, K.C. 1975, \apj, 200, l9

Walker, A.R. 1990, \aj, 100, 1532

Walker, A.R. 1992, \pasp , 104, 1063 (W92)

Walker, A.R. 1994, \aj, 108, 555

Walker, A.R., \& Nemec, J.M. 1996, \aj, 112, 2026

Wehlau, A., Liller, M.H., \& Coutts Clement, C. 1978, \aj, 83, 598  (W78)

Wehlau, A., Liller, M.H., Coutts Clement, C., \& Wizinowich, P. 1982, \aj, 87, 1295  (W82)

Welch, D., \& Stetson, P.B. 1993, \aj, 105, 1813

Yi, S., Lee, Y.-W., \& Demarque, P. 1995, \apj, 411, L25

Zinn, R., \& West, M.J. 1984, \apjs, 55, 45

\clearpage

\begin{figure}
\caption{ The full CCD field, 13.6 x 13.6 arcmin, with North up and East to the left.  The
more distant RR Lyraes are identified. From a 300s V band exposure with the CTIO 0.9-m telescope.}
\end{figure}
\begin{figure}
\caption{A 100 x 100 arcsec field, with the inner RR Lyraes identified.  Otherwise as 
Figure 1.}
\end{figure}
\begin{figure}
\caption{ V magnitude lightcurves as a function of phase for the NGC 1851 RR Lyraes.}
\end{figure}
\begin{figure}
\caption{ Log Period vs. log Temperature diagram.  Overtone periods have
been fundamentalized, and corrections made for luminosity differences, as detailed in
the text.  The lower four panels show the NGC 1851, IC 4499, NGC 6362 and M68 RR Lyraes, the 
latter plotted using a reddening of E(B-V) = 0.03.  The top
panel superimposes the results for the first three of these clusters.  Note that the diagonal 
reference line is drawn to guide the eye, it is not a fit to the data.}
\end{figure}
\begin{figure}
\caption{ Color-magnitude diagram for stars measured within an annulus radii 80 and 260 
arcsec.}
\end{figure}
\begin{figure}
\caption{ Color-magnitude diagram for stars measured within an annulus radii 30 and
 260 arcsec.}
\end{figure}
\begin{figure}
\caption{olor-magnitude diagram for stars in the vicinity of the NGC 1851 Horizontal
Branch, measured within an annulus radii 80 and 260 arcsec.}
\end{figure}
\begin{figure}
\caption{olor-magnitude diagram for stars in the vicinity of the NGC 1851 Horizontal
Branch, measured within an annulus radii 30 and 260 arcsec.}
\end{figure}
\clearpage
\pagestyle{empty}

\begin{deluxetable}{rcccc}
\footnotesize
\scriptsize
\tablewidth{0pt}
\tablenum{1}
\tablecaption{Observing Conditions}
\tablehead{
\colhead{Date}   & 
\colhead{Nobs}    &
\colhead{PM?}      &
\colhead{Seeing}   &
\colhead{ColEqn}}
\startdata
1993 Aug 23 &  4  & Yes &  1\farcs7 &  A \nl
1993 Nov 29 &  17 & No  &  1\farcs7 &  A \nl
         30 &  11 & Yes &  1\farcs5 &  A \nl
     Dec  1 &  12 & Yes &  1\farcs3 &  A \nl
1994 Mar  7 &  4  & No  &  1\farcs8 &  B \nl
          8 &  6  & Yes &  1\farcs6 &  B \nl
          9 &  6  & Yes &  1\farcs5 &  B \nl
         10 &  3  & No  &  1\farcs7 &  B \nl
1994 Sep 21 &  7  & Yes &  1\farcs6 &  C \nl
         22 &  7  & No  &  2\farcs5 &  C \nl
         24 &  8  & Yes &  1\farcs3 &  C \nl
         25 &  9  & Yes &  1\farcs3 &  C \nl
1994 Nov 23 &  8  & Yes &  1\farcs3 &  D \nl
         24 &  10 & Yes &  1\farcs5 &  D \nl
         27 &  14 & No  &  2\farcs0 &  D \nl
\enddata
\end{deluxetable}
\pagestyle{empty}

\begin{deluxetable}{cccc}
\footnotesize
\scriptsize
\tablewidth{0pt}
\tablenum{2}
\tablecaption{Color Equations}
\tablehead{
\colhead{Set}   & 
\colhead{$C_V$}    &
\colhead{$C_{B-V}$}      &
\colhead{$C_{V-I}$} }
\startdata
A & 0.027  & 0.887 & 1.002 \nl
B & 0.034  & 0.899 & 1.003 \nl
C & 0.012  & 0.890 & 1.000 \nl
D & 0.012  & 0.903 & 1.004 \nl
\enddata
\tablecomments{The color equations are of the form $V =v + C_V(b-v), B-V = 
C_{B-V}(b-v), V-I =C_{V-I}(v-i).$}
\end{deluxetable}

\begin{deluxetable}{lllllllll}
\footnotesize
\scriptsize
\tablewidth{0pt}
\tablenum{3}
\tablecaption{Constant stars in annulus radii 80, 260 arcsec}
\tablehead{
\colhead{Num}   & 
\colhead{X}    &
\colhead{Y}     &
\colhead{$V$}   &
\colhead{$B-V$} &
\colhead{$V-I$} &
\colhead{$\sigma V$}  &
\colhead{$\sigma (B-V) $}  &
\colhead{$\sigma (V-I) $}  }
\startdata
   52 &    1068.8 &    1331.1 &   13.289 &    0.690 &    0.727 &  0.000 &  0.001 &  0.001 \nl
   63 &    1158.5 &     704.5 &   13.399 &    1.576 &    1.580 &  0.001 &  0.001 &  0.002 \nl
   69 &    1095.6 &    1141.8 &   13.473 &    1.549 &    1.540 &  0.001 &  0.002 &  0.001 \nl
   82 &     434.6 &    1117.9 &   13.607 &    1.482 &    1.475 &  0.001 &  0.001 &  0.002 \nl
   90 &     933.3 &    1227.7 &   13.711 &    1.439 &    1.439 &  0.001 &  0.001 &  0.001 \nl
   96 &     412.1 &     919.6 &   13.814 &    1.482 &    1.421 &  0.001 &  0.001 &  0.002 \nl
   97 &     589.5 &     841.1 &   13.807 &    1.361 &    1.380 &  0.001 &  0.001 &  0.001 \nl
  116 &    1384.7 &    1239.3 &   14.029 &    1.310 &    1.337 &  0.001 &  0.001 &  0.001 \nl
  117 &    1280.1 &     776.0 &   14.017 &    1.333 &    1.332 &  0.000 &  0.001 &  0.001 \nl
  126 &     710.3 &     822.7 &   14.135 &    1.229 &    1.258 &  0.001 &  0.001 &  0.001 \nl
  139 &     856.0 &     695.2 &   14.239 &    1.218 &    1.261 &  0.001 &  0.001 &  0.001 \nl
  140 &    1192.4 &     331.5 &   14.272 &    0.542 &    0.624 &  0.001 &  0.002 &  0.002 \nl
  150 &     753.5 &    1177.7 &   14.304 &    1.287 &    1.255 &  0.001 &  0.001 &  0.001 \nl
  159 &     669.9 &    1032.9 &   14.372 &    1.216 &    1.254 &  0.001 &  0.002 &  0.001 \nl
  163 &     597.2 &     733.4 &   14.388 &    1.184 &    1.224 &  0.001 &  0.002 &  0.001 \nl
  169 &     749.3 &    1017.7 &   14.446 &    1.176 &    1.221 &  0.001 &  0.001 &  0.001 \nl
  181 &    1153.2 &     830.8 &   14.531 &    1.086 &    1.148 &  0.001 &  0.001 &  0.001 \nl
  192 &    1183.5 &    1043.3 &   14.573 &    1.132 &    1.174 &  0.001 &  0.001 &  0.001 \nl
  211 &    1184.7 &     527.0 &   14.705 &    1.098 &    1.173 &  0.001 &  0.001 &  0.001 \nl
  212 &    1169.8 &     970.4 &   14.697 &    0.969 &    1.064 &  0.001 &  0.001 &  0.001 \nl
  227 &     949.8 &    1155.9 &   14.771 &    1.085 &    1.145 &  0.001 &  0.002 &  0.001 \nl
  231 &     867.5 &     682.8 &   14.772 &    0.993 &    1.037 &  0.001 &  0.001 &  0.001 \nl
  241 &     733.0 &     888.6 &   14.840 &    0.968 &    1.052 &  0.001 &  0.001 &  0.001 \nl
  243 &     905.4 &     604.5 &   14.836 &    0.967 &    1.049 &  0.001 &  0.001 &  0.001 \nl
  245 &     775.8 &     793.3 &   14.850 &    0.957 &    1.011 &  0.001 &  0.001 &  0.001 \nl
  246 &    1461.4 &    1023.0 &   14.838 &    1.174 &    1.136 &  0.000 &  0.001 &  0.001 \nl
  247 &     734.5 &     928.0 &   14.864 &    1.056 &    1.121 &  0.001 &  0.001 &  0.001 \nl
  249 &     329.6 &     864.1 &   14.850 &    1.069 &    1.138 &  0.001 &  0.001 &  0.001 \nl
  258 &    1219.8 &    1320.9 &   14.914 &    1.065 &    1.132 &  0.000 &  0.001 &  0.001 \nl
  263 &    1198.6 &     918.0 &   14.930 &    0.917 &    1.024 &  0.000 &  0.001 &  0.001 \nl
  264 &     608.7 &    1341.7 &   14.929 &    1.073 &    1.142 &  0.000 &  0.001 &  0.001 \nl
  267 &    1477.2 &    1048.5 &   14.923 &    0.835 &    0.889 &  0.000 &  0.001 &  0.001 \nl
  288 &    1275.5 &    1193.2 &   15.022 &    1.048 &    1.119 &  0.001 &  0.002 &  0.001 \nl
  311 &    1207.1 &     798.4 &   15.104 &    1.089 &    1.108 &  0.000 &  0.001 &  0.001 \nl
  318 &    1048.0 &    1134.5 &   15.125 &    1.016 &    1.091 &  0.001 &  0.002 &  0.001 \nl
  334 &    1169.1 &     953.3 &   15.211 &    0.991 &    1.081 &  0.001 &  0.002 &  0.002 \nl
  339 &     893.0 &     729.5 &   15.190 &    1.104 &    1.065 &  0.001 &  0.001 &  0.001 \nl
  363 &    1150.7 &     756.1 &   15.300 &    0.604 &    0.747 &  0.001 &  0.001 &  0.001 \nl
  367 &     794.1 &     797.3 &   15.303 &    0.994 &    1.077 &  0.001 &  0.002 &  0.001 \nl
  377 &    1327.0 &     558.3 &   15.313 &    0.997 &    1.093 &  0.001 &  0.001 &  0.001 \nl
  395 &     852.0 &    1098.6 &   15.385 &    0.961 &    1.049 &  0.001 &  0.001 &  0.001 \nl
  396 &     693.7 &     998.2 &   15.388 &    0.779 &    0.888 &  0.001 &  0.001 &  0.001 \nl
  400 &    1079.9 &    1106.7 &   15.418 &    0.940 &    1.034 &  0.001 &  0.002 &  0.002 \nl
  407 &     554.8 &     472.6 &   15.416 &    0.843 &    0.967 &  0.001 &  0.001 &  0.001 \nl
  412 &    1516.7 &     847.7 &   15.420 &    0.991 &    1.069 &  0.000 &  0.001 &  0.001 \nl
  427 &    1346.1 &     879.3 &   15.471 &    0.976 &    1.048 &  0.001 &  0.001 &  0.001 \nl
  434 &     949.1 &    1174.2 &   15.489 &    0.960 &    1.042 &  0.001 &  0.001 &  0.001 \nl
\enddata
\tablecomments{Table 3 is presented in its entirety in the electronic edition of The
Astronomical Journal.  A portion is shown here for guidance regarding its form and content.}
\end{deluxetable}

\begin{deluxetable}{cccccccccc}
\footnotesize
\scriptsize
\tablewidth{0pt}
\tablenum{4}
\tablecaption{Photometry for NGC 1851-V1}
\tablehead{
\colhead{$HJD_V$}   & 
\colhead{phase}    &
\colhead{$\delta HJD_B$}  &
\colhead{$\delta HJD_I$}   &
\colhead{$V$       }   &
\colhead{$B$       }   &
\colhead{$I$       }   &
\colhead{$\sigma_V$}   &
\colhead{$\sigma_B$}   &
\colhead{$\sigma_I$}   }
\startdata
  9223.8330 &   0.434 &  0.0064 &  0.0129 &  16.014 &  16.044 &  15.384 &   0.007 &   0.016 &   0.016 \nl
  9223.8506 &   0.469 &  0.0064 &  0.0128 &  15.395 &  15.409 &  15.194 &   0.005 &   0.013 &   0.016 \nl
  9223.8691 &   0.504 &  0.0066 &  0.0129 &  15.310 &  15.432 &  15.225 &   0.005 &   0.013 &   0.011 \nl
  9223.8867 &   0.537 &  0.0064 &  0.0128 &  15.407 &  15.566 &  15.312 &   0.005 &   0.006 &   0.010 \nl
  9321.7930 &   0.609 &  0.0065 &  0.0130 &  15.609 &  15.831 &  15.415 &   0.003 &   0.012 &   0.007 \nl
  9321.8105 &   0.643 &  0.0096 &  0.0181 &  15.700 &  15.971 &  15.470 &   0.004 &   0.004 &   0.026 \nl
  9321.8428 &   0.705 &  0.0064 &  0.0130 &  15.867 &  16.160 &  15.520 &   0.004 &   0.007 &   0.005 \nl
  9322.5918 &   0.144 &  0.0064 &  0.0130 &  16.418 &  16.833 &  15.872 &   0.005 &   0.006 &   0.006 \nl
  9322.6113 &   0.182 &  0.0067 &  0.0130 &  16.449 &  16.859 &  15.883 &   0.006 &   0.028 &   0.004 \nl
  9322.6289 &   0.217 &  0.0063 &  0.0128 &  16.462 &  16.877 &  15.882 &   0.005 &   0.038 &   0.006 \nl
  9322.6465 &   0.250 &  0.0072 &  0.0137 &  16.434 &  16.800 &  15.892 &   0.006 &   0.008 &   0.017 \nl
  9322.6660 &   0.287 &  0.0063 &  0.0162 &  16.448 &  16.844 &  15.961 &   0.006 &   0.006 &   0.009 \nl
  9322.6865 &   0.326 &  0.0063 &  0.0131 &  16.529 &  16.952 &  16.003 &   0.006 &   0.005 &   0.005 \nl
  9322.7061 &   0.363 &  0.0083 &  0.0147 &  16.566 &  16.962 &  16.001 &   0.005 &   0.010 &   0.009 \nl
  9322.7314 &   0.414 &  0.0082 &  0.0147 &  16.407 &  16.549 &  15.602 &   0.005 &   0.014 &   0.005 \nl
  9322.7510 &   0.449 &  0.0064 &  0.0128 &  15.789 &  15.762 &  15.252 &   0.005 &   0.007 &   0.070 \nl
  9322.7686 &   0.484 &  0.0063 &  0.0127 &  15.291 &  15.364 &  15.183 &   0.005 &   0.009 &   0.032 \nl
  9322.7861 &   0.518 &  0.0063 &  0.0131 &  15.303 &  15.447 &  15.239 &   0.005 &   0.007 &   0.010 \nl
  9322.8037 &   0.553 &  0.0064 &  0.0127 &  15.427 &  15.595 &  15.319 &   0.006 &   0.007 &   0.004 \nl
  9322.8213 &   0.586 &  0.0064 &  0.0127 &  15.544 &  15.739 &  15.380 &   0.005 &   0.006 &   0.004 \nl
  9323.5791 &   0.041 &  0.0063 &  0.0127 &  16.413 &  16.855 &  15.837 &   0.005 &   0.032 &   0.005 \nl
  9323.5967 &   0.074 &  0.0065 &  0.0130 &  16.422 &  16.830 &  15.840 &   0.005 &   0.004 &   0.013 \nl
  9323.6152 &   0.109 &  0.0064 &  0.0134 &  16.410 &  16.829 &  15.846 &   0.006 &   0.004 &   0.005 \nl
  9323.6338 &   0.144 &  0.0063 &  0.0128 &  16.412 &  16.832 &  15.864 &   0.007 &   0.054 &   0.008 \nl
  9323.6514 &   0.180 &  0.0067 &  0.0132 &  16.428 &  16.851 &  15.877 &   0.008 &   0.015 &   0.021 \nl
  9323.6689 &   0.213 &  0.0067 &  0.0133 &  16.440 &  16.857 &  15.875 &   0.008 &   0.013 &   0.004 \nl
  9323.6885 &   0.250 &  0.0064 &  0.0129 &  16.406 &  16.818 &  15.878 &   0.007 &   0.009 &   0.003 \nl
  9323.7129 &   0.297 &  0.0063 &  0.0127 &  16.473 &  16.913 &  15.964 &   0.005 &   0.005 &   0.104 \nl
  9323.7314 &   0.332 &  0.0085 &  0.0155 &  16.543 &  16.971 &  16.004 &   0.005 &   0.011 &   0.043 \nl
  9323.7510 &   0.371 &  0.0071 &  0.0136 &  16.565 &  16.956 &  15.991 &   0.007 &   0.007 &   0.022 \nl
  9323.7695 &   0.406 &  0.0064 &  0.0128 &  16.464 &  16.723 &  15.764 &   0.007 &   0.006 &   0.044 \nl
  9323.7871 &   0.440 &  0.0063 &  0.0127 &  15.972 &  15.999 &  15.362 &   0.009 &   0.006 &   0.009 \nl
  9323.8037 &   0.473 &  0.0065 &  0.0130 &  15.395 &  15.397 &  15.180 &   0.006 &   0.006 &   0.012 \nl
  9323.8271 &   0.518 &  0.0072 &  0.0138 &  15.315 &  15.465 &  15.265 &   0.004 &   0.022 &   0.025 \nl
  9419.5537 &   0.402 &  0.0066 &  0.0134 &  16.538 &  16.846 &  15.836 &   0.006 &   0.011 &   0.004 \nl
  9419.5723 &   0.438 &  0.0066 &  0.0129 &  16.130 &  16.168 &  15.450 &   0.005 &   0.010 &   0.007 \nl
  9419.5908 &   0.473 &  0.0069 &  0.0135 &  15.486 &  15.454 &  15.189 &   0.004 &   0.013 &   0.004 \nl
  9419.6084 &   0.508 &  0.0065 &  0.0126 &  15.304 &  15.434 &  15.227 &   0.003 &   0.014 &   0.005 \nl
  9419.6270 &   0.543 &  0.0064 &  0.0127 &  15.404 &  15.582 &  15.312 &   0.003 &   0.013 &   0.060 \nl
  9420.5264 &   0.270 &  0.0083 &  0.0152 &  16.457 &  16.887 &  15.934 &   0.005 &   0.011 &   0.009 \nl
  9420.5488 &   0.315 &  0.0078 &  0.0144 &  16.526 &  17.007 &  16.018 &   0.006 &   0.022 &   0.011 \nl
  9420.5869 &   0.389 &  0.0063 &  0.0126 &  16.578 &  16.962 &  15.958 &   0.013 &   0.008 &   0.008 \nl
  9420.6055 &   0.422 &  0.0067 &  0.0130 &  16.391 &  16.549 &  15.600 &   0.007 &   0.008 &   0.019 \nl
  9420.6230 &   0.457 &  0.0067 &  0.0130 &  15.789 &  15.771 &  15.256 &   0.005 &   0.009 &   0.013 \nl
  9420.6406 &   0.490 &  0.0123 &  0.0126 &  15.334 &  15.455 &  15.209 &   0.005 &   0.027 &   0.005 \nl
  9421.5146 &   0.168 &  0.0065 &  0.0127 &  16.457 &  16.903 &  15.895 &   0.005 &   0.013 &   0.071 \nl
  9421.5322 &   0.203 &  0.0063 &  0.0127 &  16.486 &  16.923 &  99.999 &   0.005 &   0.013 &   9.999 \nl
  9421.5674 &   0.270 &  0.0065 &  0.0129 &  16.462 &  16.895 &  15.926 &   0.005 &   0.004 &   0.022 \nl
  9421.5850 &   0.305 &  0.0063 &  0.0133 &  16.513 &  16.960 &  15.988 &   0.005 &   0.014 &   0.012 \nl
  9421.6094 &   0.352 &  0.0067 &  0.0131 &  16.579 &  17.034 &  16.029 &   0.005 &   0.010 &   0.111 \nl
  9421.6270 &   0.385 &  0.0064 &  0.0127 &  16.576 &  16.981 &  15.992 &   0.005 &   0.005 &   0.007 \nl
  9422.5850 &   0.225 &  0.0031 &  0.0115 &  16.502 &  16.936 &  15.904 &   0.004 &   0.005 &   0.004 \nl
  9422.6045 &   0.262 &  0.0032 &  0.0115 &  16.474 &  16.895 &  15.907 &   0.005 &   0.106 &   0.005 \nl
  9422.6221 &   0.297 &  0.0032 &  0.0115 &  16.473 &  16.952 &  15.961 &   0.007 &   0.013 &   0.009 \nl
  9617.7559 &   0.137 &  0.0017 &  0.0045 &  16.418 &  16.844 &  15.864 &   0.005 &   0.017 &   0.009 \nl
  9617.7686 &   0.162 &  0.0107 &  0.0136 &  16.439 &  16.875 &  15.874 &   0.006 &   0.007 &   0.006 \nl
  9617.7910 &   0.205 &  0.0073 &  0.0100 &  16.473 &  16.904 &  15.892 &   0.005 &   0.017 &   0.005 \nl
  9617.8086 &   0.238 &  0.0063 &  0.0184 &  16.465 &  16.875 &  15.890 &   0.005 &   0.006 &   0.014 \nl
  9617.8359 &   0.291 &  0.0070 &  0.0133 &  16.442 &  16.874 &  15.913 &   0.006 &   0.016 &   0.054 \nl
  9619.7461 &   0.961 &  0.0064 &  0.0128 &  16.334 &  16.767 &  15.775 &   0.005 &   0.010 &   0.003 \nl
  9619.7637 &   0.994 &  0.0063 &  0.0127 &  16.360 &  16.815 &  15.790 &   0.005 &   0.018 &   0.009 \nl
  9619.8271 &   0.117 &  0.0067 &  0.0129 &  16.413 &  16.842 &  15.838 &   0.006 &   0.023 &   0.007 \nl
  9619.8516 &   0.162 &  0.0063 &  0.0126 &  16.442 &  16.859 &  15.872 &   0.005 &   0.006 &   0.006 \nl
  9619.8701 &   0.197 &  0.0062 &  0.0140 &  16.464 &  16.897 &  15.883 &   0.005 &   0.005 &   0.007 \nl
  9619.8887 &   0.234 &  0.0064 &  0.0126 &  16.466 &  16.877 &  15.879 &   0.005 &   0.011 &   0.026 \nl
  9620.7568 &   0.902 &  0.0062 &  0.0125 &  16.245 &  16.653 &  15.726 &   0.005 &   0.004 &   0.004 \nl
  9620.7744 &   0.935 &  0.0063 &  0.0126 &  16.308 &  16.729 &  15.761 &   0.004 &   0.043 &   0.033 \nl
  9620.7910 &   0.969 &  0.0063 &  0.0127 &  16.350 &  16.774 &  15.785 &   0.005 &   0.005 &   0.060 \nl
  9620.8125 &   0.008 &  0.0063 &  0.0126 &  16.383 &  16.823 &  15.803 &   0.005 &   0.014 &   0.018 \nl
  9620.8301 &   0.041 &  0.0063 &  0.0136 &  16.410 &  16.841 &  15.829 &   0.004 &   0.005 &   0.022 \nl
  9620.8477 &   0.076 &  0.0064 &  0.0129 &  16.427 &  16.842 &  15.836 &   0.004 &   0.029 &   0.015 \nl
  9620.8652 &   0.111 &  0.0063 &  0.0127 &  16.423 &  16.839 &  15.840 &   0.005 &   0.004 &   0.007 \nl
  9620.8828 &   0.143 &  0.0063 &  0.0127 &  16.429 &  16.845 &  15.851 &   0.004 &   0.005 &   0.006 \nl
  9621.7314 &   0.773 &  0.0066 &  0.0129 &  16.008 &  16.361 &  15.598 &   0.004 &   0.013 &   0.040 \nl
  9621.7490 &   0.809 &  0.0107 &  0.0170 &  16.080 &  16.455 &  15.608 &   0.005 &   0.043 &   0.020 \nl
  9621.7705 &   0.850 &  0.0062 &  0.0126 &  16.141 &  16.523 &  15.642 &   0.007 &   0.007 &   0.026 \nl
  9621.7881 &   0.883 &  0.0063 &  0.0127 &  16.209 &  16.603 &  15.688 &   0.005 &   0.005 &   0.009 \nl
  9621.8057 &   0.918 &  0.0064 &  0.0128 &  16.269 &  16.688 &  15.729 &   0.004 &   0.012 &   0.009 \nl
  9621.8232 &   0.951 &  0.0063 &  0.0126 &  16.318 &  16.751 &  15.766 &   0.005 &   0.007 &   0.019 \nl
  9621.8408 &   0.984 &  0.0062 &  0.0123 &  16.366 &  16.797 &  15.793 &   0.006 &   0.005 &   0.017 \nl
  9621.8574 &   0.016 &  0.0062 &  0.0126 &  16.399 &  16.837 &  15.808 &   0.006 &   0.005 &   0.013 \nl
  9621.8740 &   0.049 &  0.0063 &  0.0126 &  16.409 &  16.846 &  15.822 &   0.007 &   0.022 &   0.008 \nl
  9680.6006 &   0.857 &  0.0058 &  0.0116 &  16.144 &  16.551 &  15.668 &   0.004 &   0.008 &   0.029 \nl
  9680.6162 &   0.887 &  0.0058 &  0.0117 &  16.194 &  16.615 &  15.703 &   0.003 &   0.010 &   0.016 \nl
  9680.6318 &   0.918 &  0.0060 &  0.0118 &  16.245 &  16.680 &  15.730 &   0.004 &   0.011 &   0.006 \nl
  9680.6475 &   0.949 &  0.0058 &  0.0116 &  16.293 &  16.736 &  15.752 &   0.004 &   0.066 &   0.055 \nl
  9680.6641 &   0.979 &  0.0058 &  0.0116 &  16.328 &  16.789 &  15.755 &   0.005 &   0.009 &   0.023 \nl
  9680.7285 &   0.103 &  0.0059 &  0.0118 &  16.393 &  16.838 &  15.838 &   0.004 &   0.004 &   0.009 \nl
  9680.7461 &   0.137 &  0.0065 &  0.0123 &  16.402 &  16.846 &  15.850 &   0.005 &   0.004 &   0.012 \nl
  9680.7617 &   0.168 &  0.0060 &  0.0118 &  16.412 &  16.862 &  15.861 &   0.005 &   0.010 &   0.011 \nl
  9681.6016 &   0.781 &  0.0058 &  0.0118 &  16.023 &  16.376 &  15.594 &   0.004 &   0.005 &   0.006 \nl
  9681.6182 &   0.812 &  0.0058 &  0.0118 &  16.074 &  16.446 &  15.627 &   0.004 &   0.004 &   0.023 \nl
  9681.6338 &   0.844 &  0.0058 &  0.0118 &  16.131 &  16.519 &  15.657 &   0.004 &   0.006 &   0.085 \nl
  9681.6514 &   0.877 &  0.0058 &  0.0116 &  16.191 &  16.601 &  15.693 &   0.004 &   0.006 &   0.036 \nl
  9681.6670 &   0.906 &  0.0058 &  0.0117 &  16.239 &  16.665 &  15.721 &   0.004 &   0.010 &   0.011 \nl
  9681.6826 &   0.938 &  0.0058 &  0.0117 &  16.295 &  16.730 &  15.764 &   0.004 &   0.037 &   0.009 \nl
  9681.6992 &   0.967 &  0.0058 &  0.0117 &  16.344 &  16.784 &  15.777 &   0.005 &   0.016 &   0.006 \nl
  9681.7148 &   0.998 &  0.0059 &  0.0119 &  16.376 &  16.826 &  15.810 &   0.005 &   0.022 &   0.018 \nl
  9681.7402 &   0.047 &  0.0058 &  0.0119 &  16.404 &  16.852 &  15.832 &   0.004 &   0.004 &   0.007 \nl
  9681.7559 &   0.078 &  0.0058 &  0.0119 &  16.421 &  16.866 &  15.844 &   0.004 &   0.028 &   0.007 \nl
  9684.5537 &   0.451 &  0.0066 &  0.0128 &  16.022 &  16.088 &  15.400 &   0.003 &   0.016 &   0.049 \nl
  9684.5703 &   0.484 &  0.0061 &  0.0122 &  15.449 &  15.477 &  15.190 &   0.002 &   0.006 &   0.015 \nl
  9684.5869 &   0.516 &  0.0058 &  0.0116 &  15.289 &  15.440 &  15.215 &   0.002 &   0.037 &   0.020 \nl
  9684.6025 &   0.545 &  0.0058 &  0.0116 &  15.368 &  15.552 &  15.279 &   0.002 &   0.020 &   0.028 \nl
  9684.6270 &   0.592 &  0.0062 &  0.0143 &  15.526 &  15.745 &  15.373 &   0.002 &   0.004 &   0.015 \nl
  9684.6484 &   0.635 &  0.0062 &  0.0125 &  15.658 &  15.915 &  15.432 &   0.003 &   0.003 &   0.012 \nl
  9684.6660 &   0.666 &  0.0060 &  0.0119 &  15.750 &  16.029 &  15.467 &   0.003 &   0.002 &   0.009 \nl
  9684.6816 &   0.697 &  0.0059 &  0.0131 &  15.832 &  16.137 &  15.512 &   0.003 &   0.010 &   0.046 \nl
  9684.7021 &   0.736 &  0.0058 &  0.0119 &  15.925 &  16.261 &  15.556 &   0.003 &   0.004 &   0.005 \nl
  9684.7188 &   0.768 &  0.0061 &  0.0123 &  15.987 &  16.348 &  15.589 &   0.003 &   0.004 &   0.010 \nl
  9684.7354 &   0.801 &  0.0058 &  0.0119 &  16.047 &  16.423 &  15.613 &   0.003 &   0.023 &   0.038 \nl
  9684.7510 &   0.830 &  0.0061 &  0.0122 &  16.103 &  16.497 &  15.647 &   0.003 &   0.005 &   0.017 \nl
  9684.7686 &   0.865 &  0.0058 &  0.0120 &  16.166 &  16.583 &  15.687 &   0.003 &   0.006 &   0.015 \nl
\enddata
\tablecomments{Table 4 is presented in its entirety in the electronic edition of The Astronomical
Journal.  A portion is shown here for guidance regarding its form and content.}
\end{deluxetable}
\pagestyle{empty}

\begin{deluxetable}{lllllllllll}
\footnotesize
\scriptsize
\tablewidth{0pt}
\tablenum{5}
\tablecaption{Periods and mean magnitudes for the RR Lyrae variables}
\tablehead{
\colhead{Var}   & 
\colhead{Num}    &
\colhead{Period(d)} &
\colhead{$<V>$}       &
\colhead{$<B>$}    &
\colhead{$<I>$}   &
\colhead{$(B-V)_m$  }     &
\colhead{$(V-I)_m$  }   &
\colhead{$A_V$  }       &
\colhead{$A_B$  }       &
\colhead{$A_I$  }  }
\startdata
    1 &   628 &  0.520578 &   16.050 &   16.340 &   15.646 &  0.339 &  0.451 &  1.37 &  1.70 &  0.87 \nl
    3 &   586 &  0.322152 &   16.054 &   16.307 &   15.712 &  0.261 &  0.352 &  0.48 &  0.64 &  0.28 \nl
    4 &   703 &  0.585110 &   16.127 &   16.489 &   15.641 &  0.393 &  0.571 &  0.98 &  1.28 &  0.59 \nl
    5 &   702 &  0.587860 &   16.039 &   16.404 &   15.554 &  0.386 &  0.499 &  0.67 &  0.88 &  0.45 \nl
    6 &   661 &  0.606623 &   16.092 &   16.473 &   15.589 &  0.404 &  0.521 &  0.88 &  1.13 &  0.55 \nl
    7 &   701 &  0.585185 &   16.044 &   16.380 &   15.582 &  0.376 &  0.492 &  1.11 &  1.45 &  0.71 \nl
    8 &   603 &  0.511000 &   16.072 &   16.334 &   15.696 &  0.320 &  0.429 &  1.26 &  1.60 &  0.73 \nl
   10 &   684 &  0.499750 &   16.119 &   16.454 &   15.719 &  0.360 &  0.419 &  0.62 &  0.82 &  0.34 \nl
   11 &   486 &  0.667930 &   15.937 &   16.321 &   15.430 &  0.405 &  0.523 &  0.82 &  1.06 &  0.51 \nl
   12 &   673 &  0.575960 &   16.122 &   16.488 &   15.632 &  0.392 &  0.511 &  0.96 &  1.24 &  0.61 \nl
   13 &   626 &  0.282540 &   16.118 &   16.343 &   15.845 &  0.233 &  0.283 &  0.59 &  0.68 &  0.36 \nl
   14 &   418 &  0.594010 &   15.429 &   15.919 &   14.762 &  0.484 &  0.672 &  0.80 &  0.91 &  0.40 \nl
   15 &   749 &  0.541320 &   16.016 &   16.311 &   15.612 &  0.338 &  0.446 &  1.30 &  1.63 &  0.89 \nl
   16 &   575 &  0.488690 &   16.107 &   16.394 &   15.712 &  0.333 &  0.438 &  1.19 &  1.49 &  0.75 \nl
   17 &   612 &  0.700307 &   16.101 &   16.419 &   15.454 &  0.427 &  0.554 &  0.53 &  0.68 &  0.35 \nl
   18 &   544 &  0.272091 &   16.067 &   16.295 &   15.784 &  0.237 &  0.293 &  0.51 &  0.65 &  0.31 \nl
   19 &   505 &  0.405161 &   15.851 &   16.148 &   15.488 &  0.302 &  0.364 &  0.47 &  0.56 &  0.40 \nl
   20 &   517 &  0.559470 &   15.935 &   16.288 &   15.457 &  0.376 &  0.493 &  0.74 &  1.06 &  0.49 \nl
   21 &   647 &  0.268521 &   16.111 &   16.328 &   15.851 &  0.226 &  0.269 &  0.50 &  0.63 &  0.30 \nl
   22 &   753 &  0.559390 &   16.084 &   16.391 &   15.562 &  0.343 &  0.539 &  0.98 &  1.20 &  0.64 \nl
   23 &   688 &  0.265830 &   16.112 &   16.324 &   15.853 &  0.214 &  0.262 &  0.26 &  0.32 &  0.17 \nl
   26 &   710 &  0.328683 &   16.111 &   16.399 &   15.746 &  0.296 &  0.374 &  0.46 &  0.59 &  0.28 \nl
   27 &   919 &  0.523230 &   16.084 &   16.412 &   15.720 &  0.375 &  0.412 &  1.02 &  1.27 &  0.70 \nl
   28 &   654 &  0.646670 &   16.082 &   16.480 &   15.540 &  0.415 &  0.556 &  0.68 &  0.92 &  0.42 \nl
   29 &   646 &  0.603530 &   15.991 &   16.379 &   15.518 &  0.402 &  0.488 &  0.82 &  0.90 &  0.55 \nl
   30 &   448 &  0.539400 &   15.868 &   16.084 &   15.306 &  0.350 &  0.450 &  0.90 &  1.30 &  0.50 \nl
   31 &   572 &  0.426653 &   15.964 &   16.378 &   15.457 &  0.426 &  0.522 &  0.61 &  0.71 &  0.40 \nl
   32 &   764 &  0.659708 &   16.119 &   16.515 &   15.605 &  0.405 &  0.519 &  0.53 &  0.67 &  0.34 \nl
   33 &   648 &  0.341231 &   16.121 &   16.383 &   15.843 &  0.273 &  0.300 &  0.53 &  0.67 &  0.26 \nl
\enddata
\end{deluxetable}
\pagestyle{empty}

\begin{deluxetable}{ccccccccccccccccccccc}
\footnotesize
\scriptsize
\tablewidth{0pt}
\tablenum{6}
\tablecaption{RR Lyrae Fourier parameters}
\tablehead{
\colhead{Var}   & 
\colhead{Num}    &
\colhead{Filt}     &
\colhead{$A_0$}    &
\colhead{$\sigma A_0$}  &
\colhead{$A_1$}   & 
\colhead{$\sigma A_1$}  &
\colhead{$R_{21}$}   &
\colhead{$\sigma R_{21}$}  &
\colhead{$R_{31}$} & 
\colhead{$\sigma R_{31}$}  &
\colhead{$R_{41}$}      &
\colhead{$\sigma R_{41}$}  &
\colhead{$\phi_1$} &
\colhead{$\sigma (\phi_1)$}  &
\colhead{$\phi_{21}$} &
\colhead{$\sigma (\phi_{21})$}  &
\colhead{$\phi_{31}$} &
\colhead{$\sigma (\phi_{31})$}  &
\colhead{$\phi_{41}$} &
\colhead{$\sigma (\phi_{41})$}  }
\startdata
 3 & 586 & V & 16.069 & 0.002 & 0.252 & 0.003 & 0.108 & 0.012 & 0.074 & 0.012 & 0.053 & 0.012 & 3.43 & 0.01 & 4.50 & 0.12 & 3.41 & 0.18 & 2.17 &  0.25 \nl
 3 & 586 & B & 16.330 & 0.002 & 0.318 & 0.003 & 0.095 & 0.009 & 0.079 & 0.008 & 0.046 & 0.008 & 3.56 & 0.01 & 4.48 & 0.10 & 3.49 & 0.12 & 2.12 &  0.20 \nl
 3 & 586 & I & 15.717 & 0.002 & 0.146 & 0.004 & 0.126 & 0.027 & 0.061 & 0.025 & 0.062 & 0.025 & 3.60 & 0.02 & 4.83 & 0.23 & 3.76 & 0.46 & 1.95 &  0.47 \nl
 13 & 626 & V & 16.139 & 0.007 & 0.281 & 0.009 & 0.180 & 0.039 & 0.033 & 0.034 & 0.048 & 0.034 & 0.25 & 0.03 & 4.81 & 0.25 & 3.06 & 1.12 & 0.77 & 0.82 \nl
 13 & 626 & B & 16.372 & 0.006 & 0.344 & 0.008 & 0.159 & 0.026 & 0.056 & 0.024 & 0.080 & 0.024 & 0.37 & 0.02 & 5.03 & 0.19 & 2.89 & 0.47 & 1.28 & 0.37 \nl
 13 & 626 & I & 15.856 & 0.010 & 0.183 & 0.014 & 0.135 & 0.087 & 0.140 & 0.090 & 0.033 & 0.078 & 0.55 & 0.08 & 5.67 & 0.71 & 3.32 & 0.75 & 6.17 & 2.59 \nl
 18 & 544 & V & 16.083 & 0.001 & 0.256 & 0.002 & 0.197 & 0.009 & 0.055 & 0.008 & 0.059 & 0.008 & 5.16 & 0.01 & 4.74 & 0.05 & 2.69 & 0.16 & 1.40 & 0.16 \nl
 18 & 544 & B & 16.320 & 0.002 & 0.322 & 0.003 & 0.189 & 0.010 & 0.049 & 0.009 & 0.053 & 0.009 & 5.29 & 0.01 & 4.68 & 0.06 & 3.03 & 0.19 & 1.58 & 0.19 \nl
 18 & 544 & I & 15.790 & 0.002 & 0.150 & 0.002 & 0.226 & 0.020 & 0.052 & 0.017 & 0.044 & 0.017 & 5.37 & 0.02 & 4.88 & 0.11 & 3.15 & 0.36 & 1.81 & 0.43 \nl
 19 & 505 & V & 15.865 & 0.008 & 0.225 & 0.012 & 0.114 & 0.058 & 0.078 & 0.056 & 0.003 & 0.052 & 6.01 & 0.05 & 4.68 & 0.56 & 5.19 & 0.82 & 0.81 & 0.88 \nl
 19 & 505 & B & 16.167 & 0.006 & 0.274 & 0.008 & 0.037 & 0.030 & 0.076 & 0.031 & 0.057 & 0.031 & 6.09 & 0.03 & 4.06 & 0.84 & 4.37 & 0.46 & 2.36 & 0.62 \nl
 19 & 505 & I & 15.501 & 0.012 & 0.170 & 0.017 & 0.145 & 0.111 & 0.162 & 0.113 & 0.105 & 0.106 & 6.23 & 0.09 & 4.38 & 0.86 & 5.87 & 0.88 & 0.40 & 1.30 \nl
 21 & 647 & V & 16.126 & 0.002 & 0.250 & 0.003 & 0.207 & 0.013 & 0.053 & 0.011 & 0.052 & 0.011 & 3.42 & 0.01 & 4.64 & 0.07 & 2.82 & 0.23 & 1.52 & 0.25 \nl
 21 & 647 & B & 16.352 & 0.002 & 0.315 & 0.002 & 0.187 & 0.009 & 0.041 & 0.008 & 0.056 & 0.007 & 3.55 & 0.01 & 4.71 & 0.05 & 3.13 & 0.20 & 1.63 & 0.16 \nl
 21 & 647 & I & 15.857 & 0.002 & 0.150 & 0.003 & 0.205 & 0.023 & 0.062 & 0.020 & 0.068 & 0.020 & 3.64 & 0.02 & 4.65 & 0.13 & 2.93 & 0.36 & 1.70 & 0.35 \nl
 23 & 688 & V & 16.117 & 0.005 & 0.124 & 0.006 & 0.081 & 0.057 & 0.018 & 0.054 & 0.029 & 0.054 & 5.67 & 0.05 & 4.98 & 0.75 & 0.82 & 2.97 & 4.60 & 2.01 \nl
 23 & 688 & B & 16.331 & 0.006 & 0.154 & 0.008 & 0.085 & 0.059 & 0.013 & 0.055 & 0.029 & 0.056 & 5.78 & 0.05 & 4.67 & 0.74 & 0.64 & 4.17 & 4.28 & 2.08 \nl
 23 & 688 & I & 15.855 & 0.003 & 0.077 & 0.004 & 0.113 & 0.063 & 0.005 & 0.058 & 0.062 & 0.059 & 5.96 & 0.06 & 4.38 & 0.61 & 5.67 & 9.99 & 4.42 & 1.15 \nl
 26 & 710 & V & 16.124 & 0.002 & 0.237 & 0.003 & 0.063 & 0.012 & 0.082 & 0.013 & 0.031 & 0.011 & 2.10 & 0.01 & 4.96 & 0.22 & 3.77 & 0.18 & 2.76 & 0.41 \nl
 26 & 710 & B & 16.420 & 0.002 & 0.301 & 0.003 & 0.082 & 0.012 & 0.082 & 0.012 & 0.043 & 0.011 & 2.25 & 0.01 & 4.72 & 0.16 & 3.89 & 0.17 & 2.65 & 0.29 \nl
 26 & 710 & I & 15.750 & 0.002 & 0.142 & 0.003 & 0.061 & 0.019 & 0.076 & 0.020 & 0.046 & 0.019 & 2.22 & 0.02 & 5.22 & 0.35 & 3.98 & 0.29 & 3.39 & 0.44 \nl
 31 & 572 & V & 15.990 & 0.012 & 0.279 & 0.017 & 0.233 & 0.072 & 0.056 & 0.062 & 9.999 & 9.999 & 3.23 & 0.06 & 4.90 & 0.38 & 2.18 & 1.24 & 5.91 & 0.24 \nl
 31 & 572 & B & 16.415 & 0.014 & 0.339 & 0.020 & 0.188 & 0.068 & 0.099 & 0.063 & 9.999 & 9.999 & 3.32 & 0.06 & 4.74 & 0.43 & 2.88 & 0.76 & 5.56 & 0.23 \nl
 31 & 572 & I & 15.469 & 0.009 & 0.173 & 0.013 & 0.292 & 0.097 & 0.098 & 0.081 & 9.999 & 9.999 & 3.17 & 0.07 & 5.14 & 0.41 & 2.94 & 0.99 & 6.18 & 0.30 \nl
 33 & 648 & V & 16.223 & 0.011 & 0.238 & 0.014 & 0.195 & 0.077 & 0.132 & 0.070 & 9.999 & 9.999 & 0.28 & 0.07 & 0.45 & 0.45 & 5.09 & 0.64 & 5.17 & 0.27 \nl
 33 & 648 & B & 16.477 & 0.010 & 0.336 & 0.014 & 0.122 & 0.044 & 0.159 & 0.046 & 9.999 & 9.999 & 0.49 & 0.04 & 5.54 & 0.46 & 4.07 & 0.38 & 4.33 & 0.18 \nl
 33 & 648 & I & 15.919 & 0.019 & 0.086 & 0.026 & 0.554 & 0.447 & 0.541 & 0.465 & 9.999 & 9.999 & 0.58 & 0.32 & 5.12 & 1.28 & 4.83 & 1.51 & 3.97 & 1.29 \nl
 1 & 628 & V & 16.124 & 0.004 & 0.452 & 0.006 & 0.478 & 0.019 & 0.341 & 0.017 & 0.215 & 0.015 & 5.48 & 0.01 & 3.91 & 0.05 & 1.90 & 0.08 & 6.13 &  0.11 \nl
 1 & 628 & B & 16.463 & 0.005 & 0.579 & 0.007 & 0.474 & 0.018 & 0.332 & 0.016 & 0.204 & 0.015 & 5.60 & 0.01 & 3.82 & 0.05 & 1.75 & 0.07 & 5.95 &  0.11 \nl
 1 & 628 & I & 15.673 & 0.002 & 0.269 & 0.003 & 0.504 & 0.018 & 0.382 & 0.016 & 0.251 & 0.015 & 5.45 & 0.01 & 4.26 & 0.05 & 2.43 & 0.07 & 0.62 &  0.10 \nl
 4 & 703 & V & 16.164 & 0.005 & 0.315 & 0.007 & 0.566 & 0.035 & 0.334 & 0.030 & 0.215 & 0.027 & 1.40 & 0.02 & 3.99 & 0.08 & 1.98 & 0.13 & 0.20 &  0.19 \nl
 4 & 703 & B & 16.557 & 0.007 & 0.433 & 0.010 & 0.523 & 0.035 & 0.344 & 0.031 & 0.204 & 0.027 & 1.56 & 0.02 & 3.76 & 0.09 & 1.67 & 0.13 & 6.25 &  0.20 \nl
 4 & 703 & I & 15.653 & 0.005 & 0.180 & 0.008 & 0.652 & 0.070 & 0.330 & 0.057 & 0.219 & 0.051 & 1.24 & 0.04 & 4.49 & 0.15 & 3.06 & 0.25 & 1.84 &  0.36 \nl
 5 & 702 & V & 16.063 & 0.010 & 0.259 & 0.015 & 0.440 & 0.082 & 0.267 & 0.070 & 0.142 & 0.064 & 0.56 & 0.05 & 4.22 & 0.23 & 2.29 & 0.36 & 6.00 &  0.56 \nl
 5 & 702 & B & 16.449 & 0.013 & 0.364 & 0.019 & 0.348 & 0.070 & 0.250 & 0.063 & 0.120 & 0.055 & 0.68 & 0.05 & 4.12 & 0.25 & 2.15 & 0.35 & 5.97 &  0.61 \nl
 5 & 702 & I & 15.564 & 0.007 & 0.147 & 0.011 & 0.679 & 0.122 & 0.221 & 0.085 & 0.142 & 0.073 & 0.55 & 0.07 & 4.42 & 0.24 & 2.38 & 0.51 & 0.83 &  0.77 \nl
 6 & 661 & V & 16.123 & 0.002 & 0.297 & 0.004 & 0.524 & 0.018 & 0.326 & 0.016 & 0.177 & 0.014 & 5.08 & 0.01 & 4.10 & 0.04 & 2.20 & 0.07 & 0.42 &  0.11 \nl
 6 & 661 & B & 16.528 & 0.003 & 0.396 & 0.005 & 0.504 & 0.017 & 0.303 & 0.015 & 0.168 & 0.014 & 5.21 & 0.01 & 3.96 & 0.04 & 2.00 & 0.07 & 0.07 &  0.11 \nl
 6 & 661 & I & 15.602 & 0.003 & 0.188 & 0.004 & 0.525 & 0.030 & 0.294 & 0.025 & 0.153 & 0.022 & 4.96 & 0.02 & 4.57 & 0.07 & 3.06 & 0.12 & 1.53 &  0.20 \nl
 7 & 701 & V & 16.095 & 0.001 & 0.363 & 0.002 & 0.520 & 0.009 & 0.333 & 0.007 & 0.233 & 0.007 & 3.43 & 0.01 & 3.99 & 0.02 & 2.03 & 0.04 & 0.07 &  0.05 \nl
 7 & 701 & B & 16.471 & 0.003 & 0.485 & 0.004 & 0.498 & 0.012 & 0.333 & 0.010 & 0.215 & 0.009 & 3.54 & 0.01 & 3.91 & 0.03 & 1.89 & 0.05 & 6.18 &  0.07 \nl
 7 & 701 & I & 15.601 & 0.001 & 0.220 & 0.002 & 0.527 & 0.015 & 0.356 & 0.012 & 0.251 & 0.012 & 3.34 & 0.01 & 4.35 & 0.04 & 2.68 & 0.06 & 1.04 &  0.08 \nl
 8 & 603 & V & 16.151 & 0.008 & 0.463 & 0.011 & 0.456 & 0.033 & 0.321 & 0.030 & 0.180 & 0.027 & 0.60 & 0.02 & 3.63 & 0.10 & 1.44 & 0.14 & 5.53 &  0.22 \nl
 8 & 603 & B & 16.471 & 0.010 & 0.609 & 0.015 & 0.452 & 0.034 & 0.294 & 0.031 & 0.164 & 0.028 & 0.67 & 0.02 & 3.72 & 0.10 & 1.42 & 0.16 & 5.55 &  0.25 \nl
 8 & 603 & I & 15.722 & 0.008 & 0.270 & 0.011 & 0.425 & 0.056 & 0.267 & 0.050 & 0.226 & 0.048 & 0.64 & 0.04 & 3.93 & 0.18 & 1.78 & 0.27 & 6.01 &  0.34 \nl
 11 & 486 & V & 15.964 & 0.003 & 0.287 & 0.004 & 0.515 & 0.021 & 0.326 & 0.019 & 0.109 & 0.016 & 3.17 & 0.02 & 4.17 & 0.07 & 2.34 & 0.10 & 0.63 & 0.20 \nl
 11 & 486 & B & 16.369 & 0.003 & 0.382 & 0.003 & 0.504 & 0.015 & 0.307 & 0.013 & 0.106 & 0.011 & 3.30 & 0.01 & 4.05 & 0.05 & 2.10 & 0.07 & 0.33 & 0.15 \nl
 11 & 486 & I & 15.441 & 0.003 & 0.182 & 0.004 & 0.480 & 0.034 & 0.317 & 0.032 & 0.102 & 0.027 & 3.10 & 0.03 & 4.46 & 0.12 & 3.01 & 0.17 & 1.79 & 0.36 \nl
 12 & 673 & V & 16.157 & 0.002 & 0.319 & 0.003 & 0.508 & 0.014 & 0.332 & 0.012 & 0.215 & 0.011 & 3.15 & 0.01 & 4.04 & 0.04 & 2.11 & 0.06 & 0.21 & 0.08 \nl
 12 & 673 & B & 16.549 & 0.003 & 0.417 & 0.003 & 0.512 & 0.012 & 0.331 & 0.011 & 0.208 & 0.010 & 3.29 & 0.01 & 3.86 & 0.03 & 1.85 & 0.05 & 6.20 & 0.08 \nl
 12 & 673 & I & 15.646 & 0.002 & 0.199 & 0.002 & 0.532 & 0.019 & 0.320 & 0.017 & 0.219 & 0.016 & 3.06 & 0.01 & 4.40 & 0.05 & 2.75 & 0.08 & 1.11 & 0.11 \nl
 14 & 418 & V & 15.431 & 0.032 & 0.262 & 0.044 & 0.347 & 0.222 & 0.148 & 0.142 & 0.302 & 0.131 & 1.00 & 0.16 & 5.25 & 0.65 & 0.33 & 1.12 & 5.96 & 0.84 \nl
 14 & 418 & B & 15.915 & 0.038 & 0.316 & 0.052 & 0.332 & 0.152 & 0.137 & 0.116 & 0.263 & 0.110 & 1.12 & 0.15 & 4.53 & 0.82 & 1.03 & 1.29 & 5.79 & 0.89 \nl
 14 & 418 & I & 14.759 & 0.029 & 0.115 & 0.040 & 0.512 & 0.538 & 0.102 & 0.254 & 0.291 & 0.250 & 1.00 & 0.33 & 5.41 & 1.03 & 5.60 & 3.21 & 0.02 & 1.86 \nl
 15 & 749 & V & 16.086 & 0.007 & 0.434 & 0.011 & 0.519 & 0.036 & 0.355 & 0.032 & 0.184 & 0.028 & 3.29 & 0.02 & 3.82 & 0.09 & 1.79 & 0.13 & 5.91 & 0.21 \nl
 15 & 749 & B & 16.424 & 0.008 & 0.545 & 0.011 & 0.519 & 0.031 & 0.348 & 0.027 & 0.200 & 0.024 & 3.40 & 0.02 & 3.72 & 0.08 & 1.73 & 0.11 & 5.72 & 0.17 \nl
 15 & 749 & I & 15.640 & 0.005 & 0.269 & 0.007 & 0.554 & 0.040 & 0.370 & 0.035 & 0.264 & 0.032 & 3.28 & 0.02 & 4.16 & 0.09 & 2.42 & 0.14 & 0.43 & 0.19 \nl
 16 & 575 & V & 16.175 & 0.005 & 0.451 & 0.007 & 0.461 & 0.021 & 0.283 & 0.019 & 0.138 & 0.017 & 2.75 & 0.01 & 3.92 & 0.06 & 1.70 & 0.10 & 5.91 & 0.17 \nl
 16 & 575 & B & 16.508 & 0.005 & 0.587 & 0.008 & 0.446 & 0.019 & 0.253 & 0.016 & 0.131 & 0.015 & 2.84 & 0.01 & 3.85 & 0.06 & 1.68 & 0.09 & 5.92 & 0.15 \nl
 16 & 575 & I & 15.737 & 0.003 & 0.271 & 0.005 & 0.473 & 0.025 & 0.297 & 0.022 & 0.149 & 0.020 & 2.76 & 0.02 & 4.20 & 0.07 & 2.15 & 0.11 & 0.22 & 0.18 \nl
 17 & 612 & V & 16.014 & 0.002 & 0.210 & 0.003 & 0.440 & 0.021 & 0.215 & 0.019 & 0.092 & 0.016 & 2.54 & 0.02 & 4.33 & 0.07 & 2.67 & 0.12 & 1.45 & 0.23 \nl
 17 & 612 & B & 16.441 & 0.002 & 0.274 & 0.003 & 0.436 & 0.016 & 0.220 & 0.014 & 0.067 & 0.012 & 2.68 & 0.01 & 4.07 & 0.05 & 2.31 & 0.09 & 1.00 & 0.21 \nl
 17 & 612 & I & 15.460 & 0.002 & 0.137 & 0.003 & 0.435 & 0.032 & 0.212 & 0.028 & 0.102 & 0.025 & 2.37 & 0.02 & 4.71 & 0.10 & 3.50 & 0.18 & 2.59 & 0.31 \nl
 20 & 517 & V & 15.962 & 0.008 & 0.281 & 0.011 & 0.431 & 0.055 & 0.273 & 0.048 & 0.156 & 0.046 & 4.99 & 0.04 & 3.89 & 0.19 & 2.39 & 0.29 & 0.34 & 0.43 \nl
 20 & 517 & B & 16.338 & 0.008 & 0.375 & 0.011 & 0.480 & 0.043 & 0.294 & 0.037 & 0.186 & 0.036 & 5.09 & 0.03 & 3.99 & 0.13 & 2.15 & 0.21 & 0.04 & 0.29 \nl
 20 & 517 & I & 15.469 & 0.009 & 0.166 & 0.011 & 0.547 & 0.104 & 0.294 & 0.087 & 0.212 & 0.086 & 4.84 & 0.08 & 4.60 & 0.29 & 2.95 & 0.49 & 2.35 & 0.64 \nl
 22 & 753 & V & 16.125 & 0.009 & 0.365 & 0.013 & 0.483 & 0.049 & 0.275 & 0.046 & 0.137 & 0.036 & 2.49 & 0.04 & 3.92 & 0.15 & 1.77 & 0.21 & 6.03 & 0.37 \nl
 22 & 753 & B & 16.468 & 0.012 & 0.497 & 0.016 & 0.440 & 0.042 & 0.200 & 0.039 & 0.082 & 0.032 & 2.63 & 0.03 & 3.82 & 0.15 & 1.70 & 0.25 & 6.22 & 0.52 \nl
 22 & 753 & I & 15.586 & 0.020 & 0.266 & 0.023 & 0.287 & 0.133 & 0.178 & 0.090 & 0.139 & 0.082 & 2.62 & 0.13 & 4.15 & 0.55 & 2.62 & 0.86 & 6.17 & 0.96 \nl
 27 & 919 & V & 16.157 & 0.014 & 0.406 & 0.019 & 0.438 & 0.046 & 0.233 & 0.050 & 0.139 & 0.030 & 6.00 & 0.05 & 3.90 & 0.23 & 1.86 & 0.31 & 5.84 & 0.47 \nl
 27 & 919 & B & 16.532 & 0.017 & 0.512 & 0.024 & 0.422 & 0.043 & 0.249 & 0.050 & 0.116 & 0.031 & 5.98 & 0.05 & 3.83 & 0.23 & 1.78 & 0.30 & 5.65 & 0.51 \nl
 27 & 919 & I & 15.745 & 0.016 & 0.232 & 0.022 & 0.480 & 0.090 & 0.365 & 0.112 & 0.238 & 0.075 & 5.80 & 0.09 & 4.08 & 0.43 & 2.30 & 0.49 & 6.25 & 0.68 \nl
 28 & 654 & V & 16.104 & 0.006 & 0.257 & 0.009 & 0.475 & 0.051 & 0.264 & 0.043 & 0.092 & 0.038 & 3.09 & 0.04 & 4.13 & 0.14 & 2.27 & 0.24 & 0.39 & 0.51 \nl
 28 & 654 & B & 16.519 & 0.009 & 0.337 & 0.012 & 0.539 & 0.056 & 0.245 & 0.045 & 0.093 & 0.039 & 3.19 & 0.04 & 4.03 & 0.14 & 2.12 & 0.26 & 0.33 & 0.54 \nl
 28 & 654 & I & 15.548 & 0.006 & 0.146 & 0.008 & 0.568 & 0.092 & 0.218 & 0.072 & 0.149 & 0.064 & 2.94 & 0.06 & 4.63 & 0.23 & 3.42 & 0.44 & 1.49 & 0.64 \nl
 29 & 646 & V & 16.011 & 0.010 & 0.288 & 0.015 & 0.534 & 0.079 & 0.234 & 0.062 & 9.999 & 9.999 & 1.36 & 0.04 & 4.06 & 0.18 & 2.16 & 0.33 & 0.84 & 0.18 \nl
 29 & 646 & B & 16.409 & 0.011 & 0.367 & 0.016 & 0.427 & 0.062 & 0.242 & 0.052 & 9.999 & 9.999 & 1.53 & 0.04 & 3.72 & 0.17 & 1.67 & 0.27 & 0.16 & 0.15 \nl
 29 & 646 & I & 15.511 & 0.013 & 0.154 & 0.019 & 0.369 & 0.154 & 0.554 & 0.181 & 9.999 & 9.999 & 1.08 & 0.11 & 5.22 & 0.55 & 3.69 & 0.53 & 1.97 & 0.44 \nl
 30 & 448 & V & 15.964 & 0.053 & 0.390 & 0.076 & 0.426 & 0.273 & 0.190 & 0.205 & 9.999 & 9.999 & 3.35 & 0.17 & 3.41 & 0.79 & 0.87 & 1.55 & 5.47 & 0.70 \nl
 30 & 448 & B & 16.203 & 0.063 & 0.396 & 0.091 & 0.520 & 0.334 & 0.511 & 0.317 & 9.999 & 9.999 & 3.29 & 0.21 & 3.89 & 0.86 & 1.63 & 1.05 & 5.70 & 0.83 \nl
 30 & 448 & I & 15.355 & 0.054 & 0.179 & 0.083 & 0.486 & 0.656 & 0.313 & 0.541 & 9.999 & 9.999 & 3.15 & 0.38 & 5.01 & 1.66 & 4.73 & 2.48 & 6.25 & 1.52 \nl
 \enddata
 \end{deluxetable}  
\pagestyle{empty}

\begin{deluxetable}{cccccc}
\scriptsize
\tablewidth{0pt}
\tablenum{7}
\tablecaption{Instability Strip Boundaries}
\tablehead{
\colhead{Cluster} &
\colhead{[Fe/H]}  &
\colhead{E(B-V)} &
\colhead{$B-V_{HBE}$} &
\colhead{$B-V_{FRE}$}  &
\colhead{$B-V_{T}$}    }
\startdata
6362 & -1.1  &  0.04  &  0.18  &  0.44   &  0.28  \nl
1851 & -1.3  &  0.02  &  0.19  &  0.41   &  0.29  \nl
4499 & -1.6  &  0.22  &  0.17  &  0.39   &  0.29  \nl
M72  & -1.6  &  0.05  &  0.18  &  0.39   &  0.28  \nl
Ret. & -1.7  &  0.03  &  0.17  &  0.40   &  0.28  \nl
2257 & -1.8  &  0.04  &  0.18  &  0.37   &  0.28  \nl
1466 & -1.8  &  0.09  &  0.16  &  0.42   &  0.27  \nl
1841 & -2.2  &  0.18  &  0.18  &  0.37   &   ..   \nl
M68  & -2.2  &  0.07  &  0.17  &  0.39   &  0.28  \nl
\enddata 
\end{deluxetable}
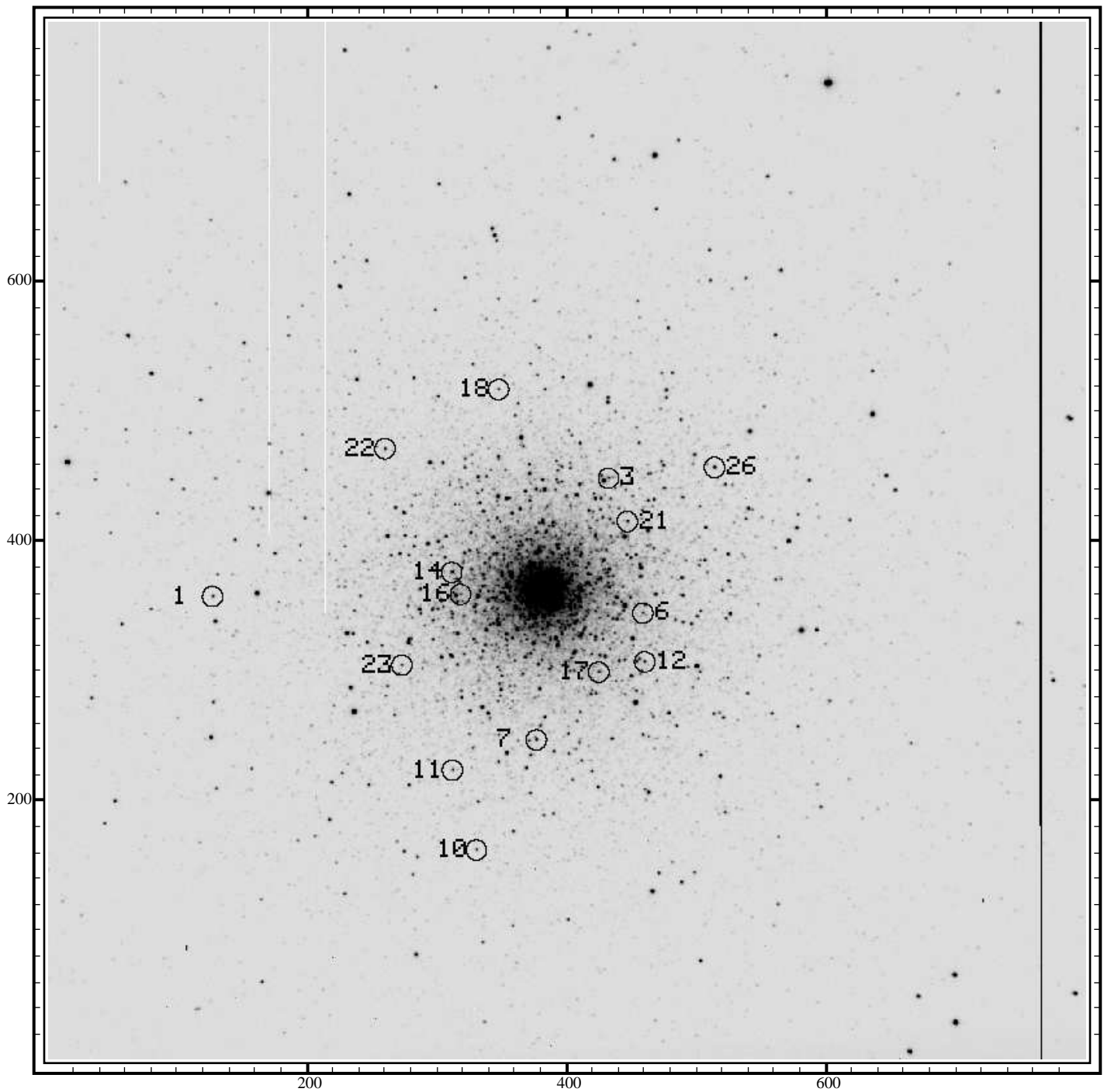
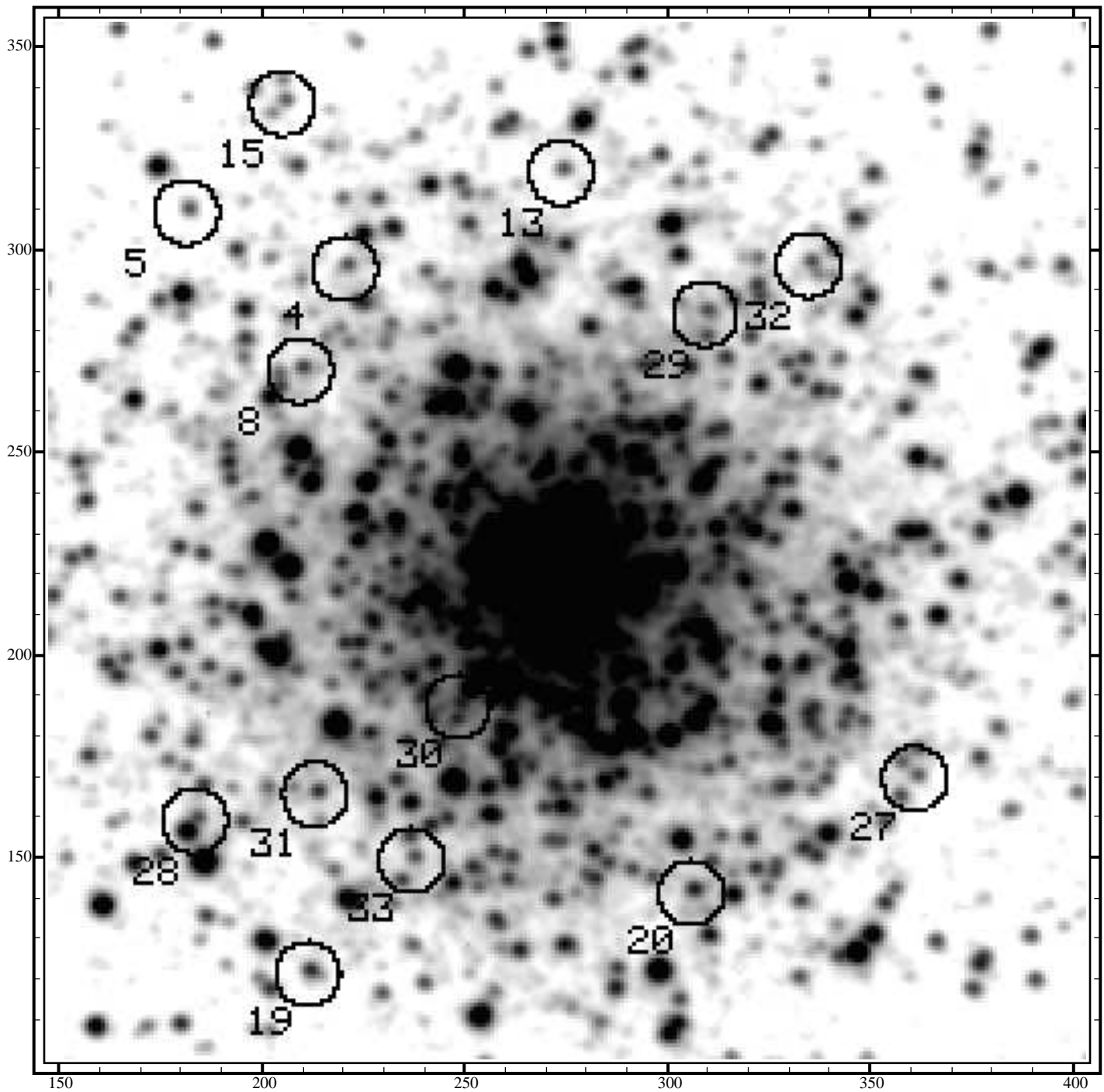
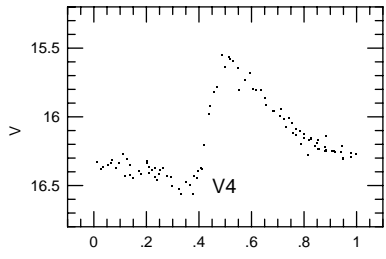
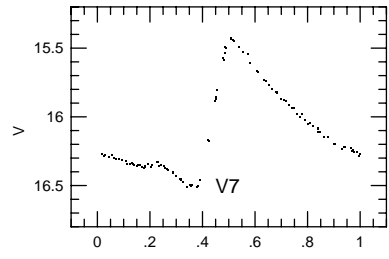
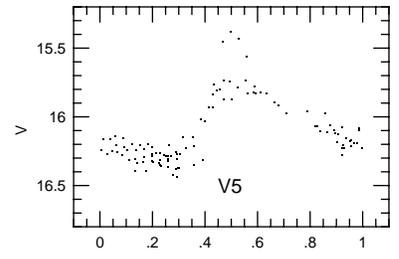
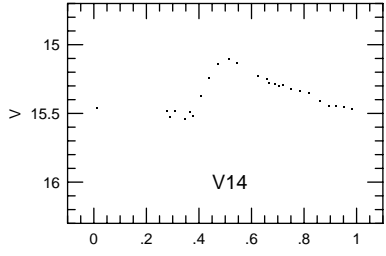
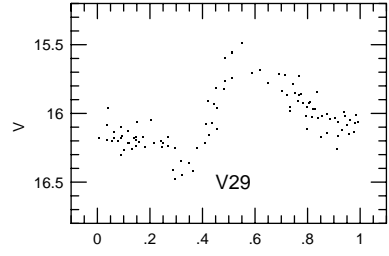
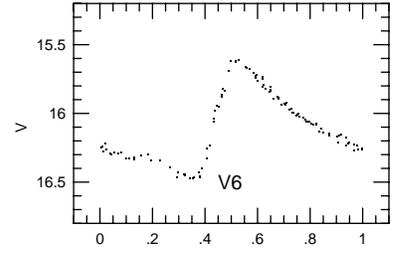
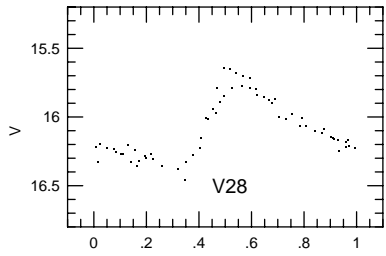
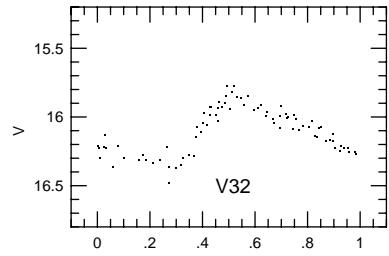
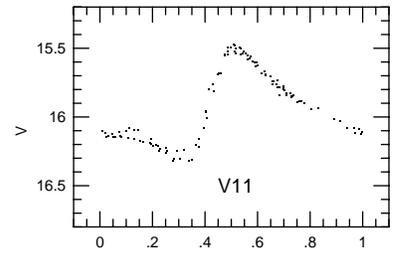
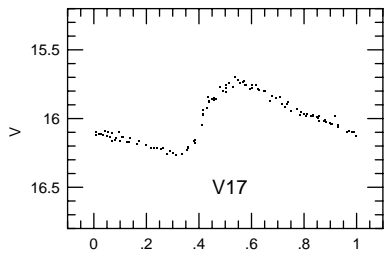
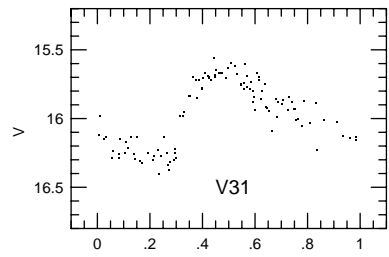
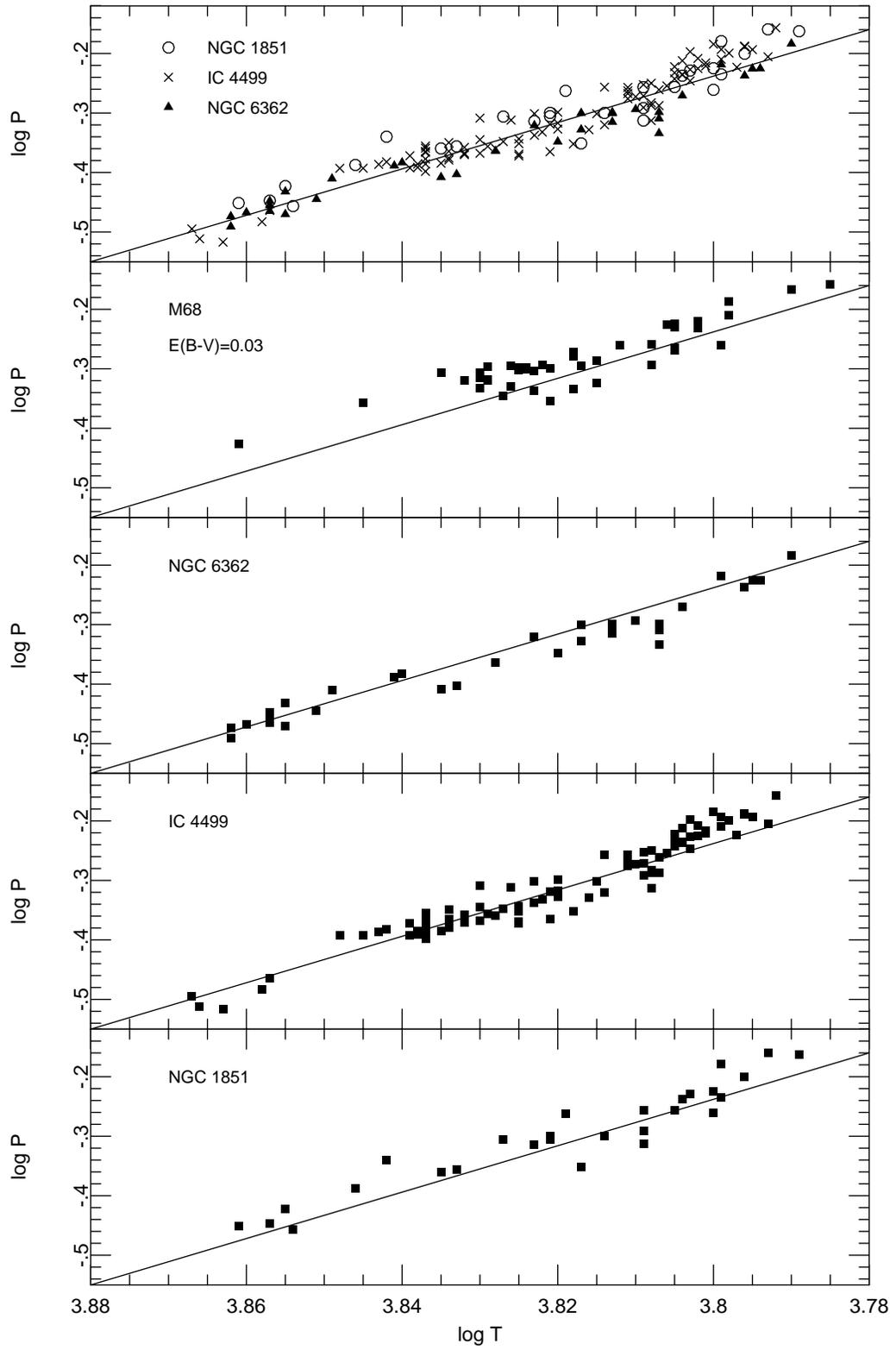
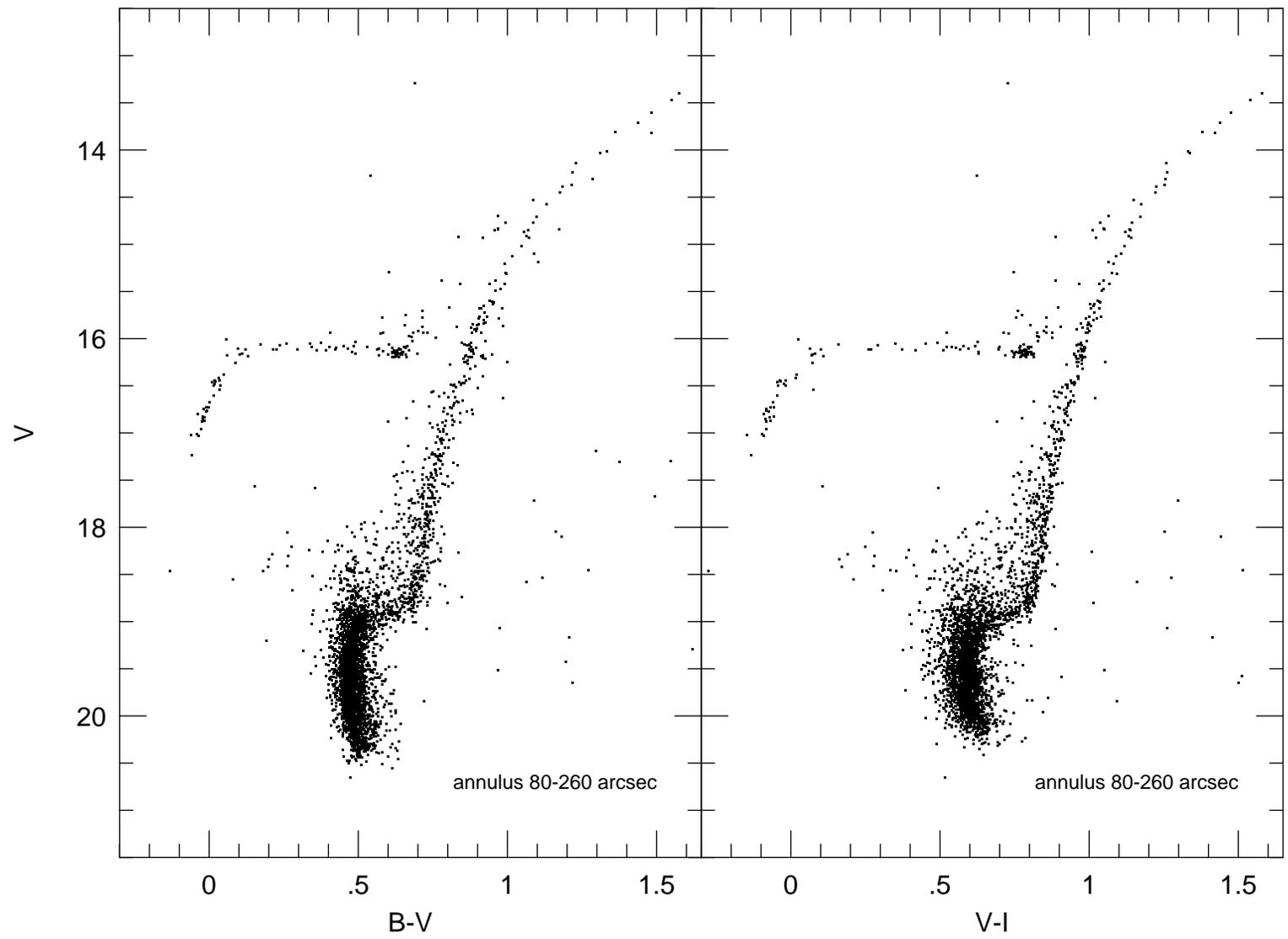
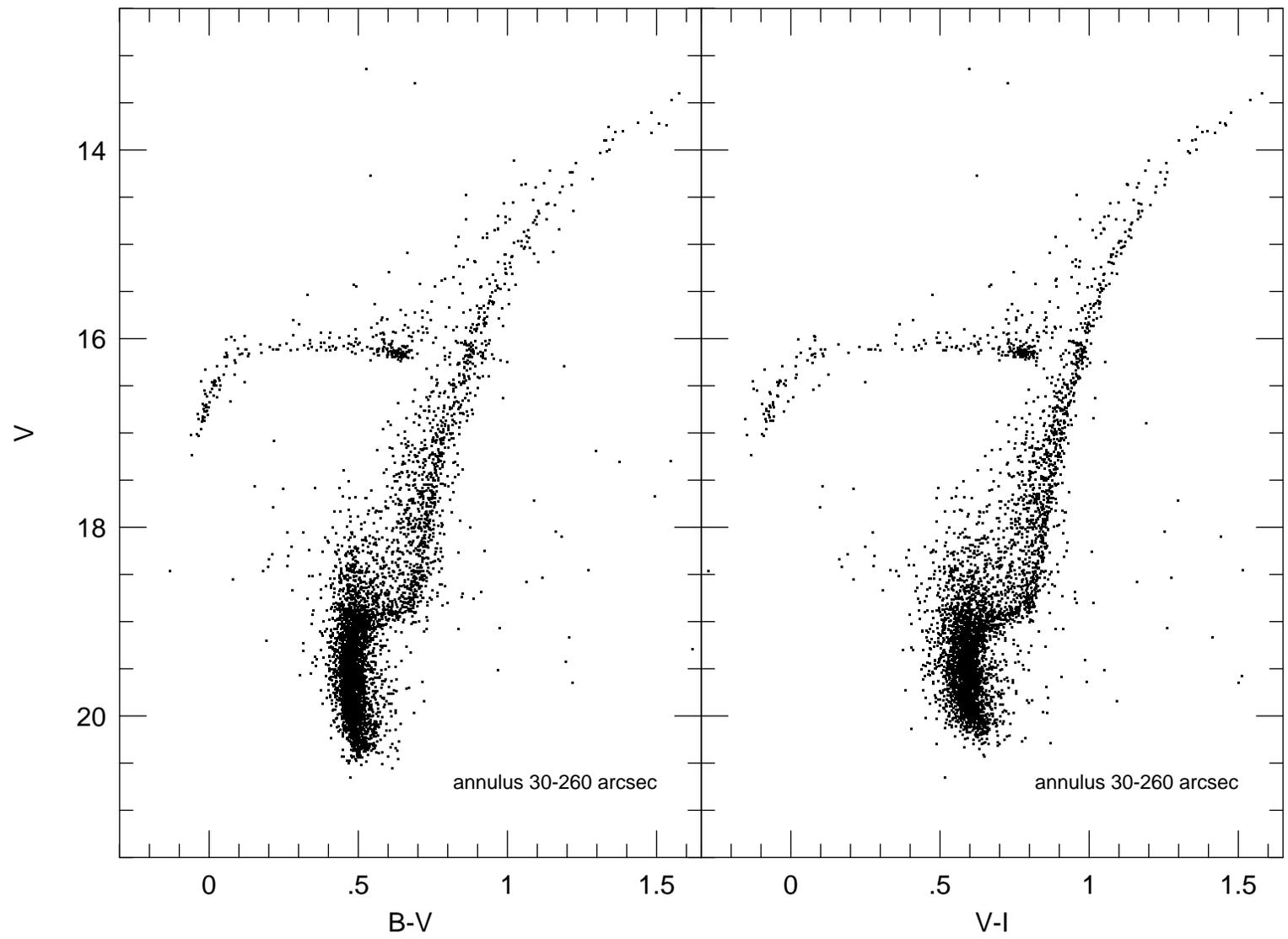
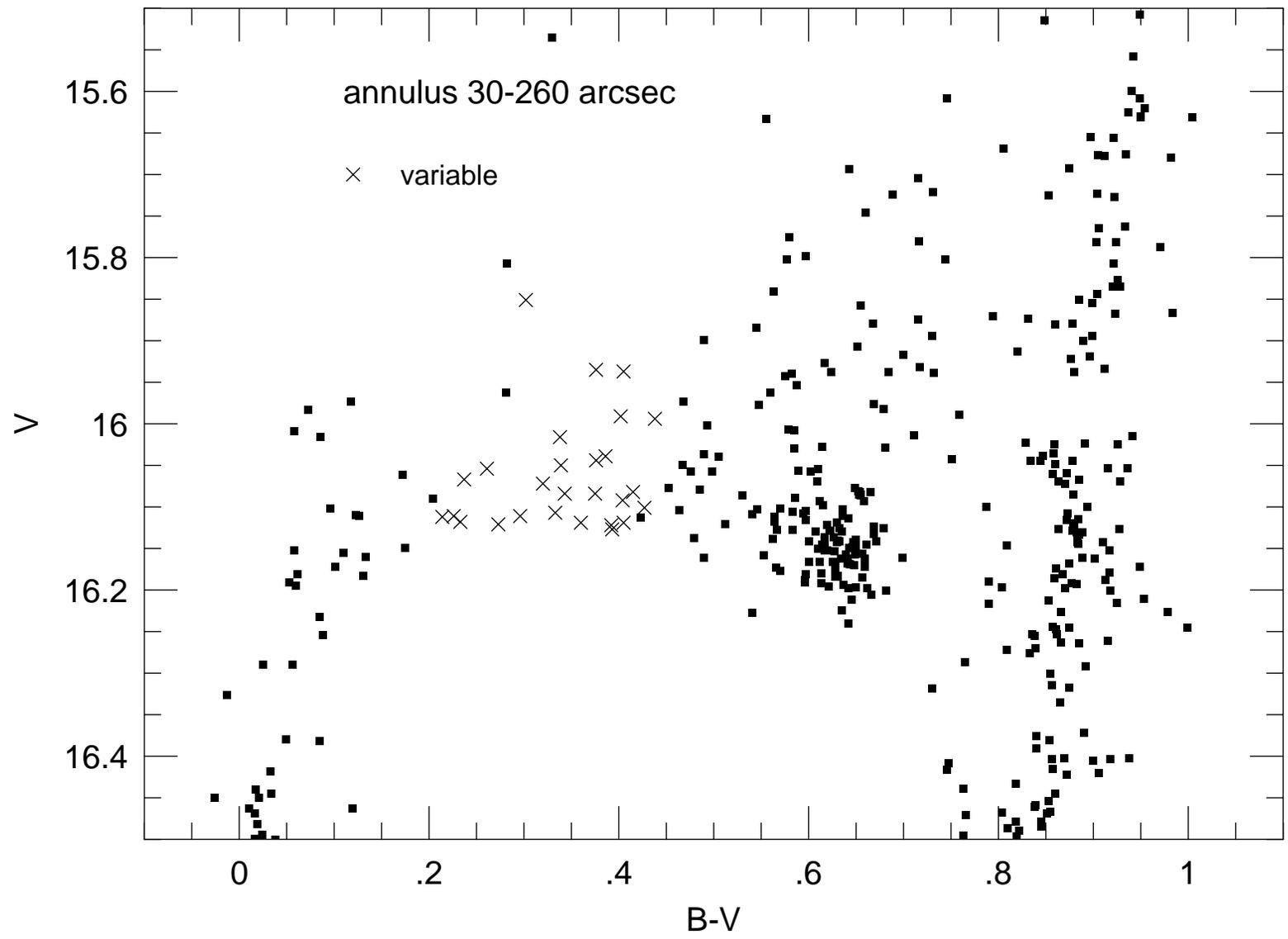
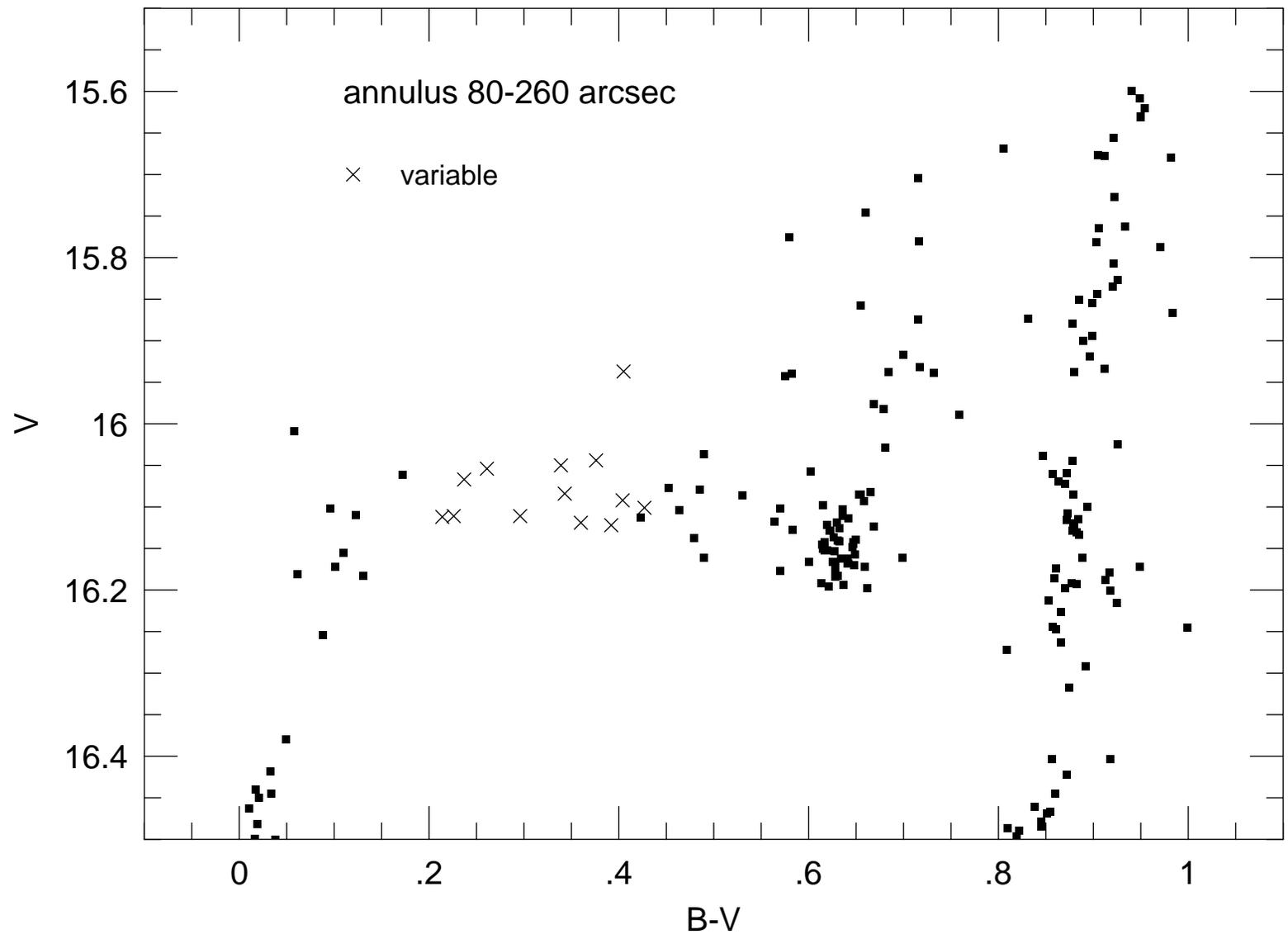
\end{document}